\documentclass{jpp}
\usepackage{graphicx}
\usepackage{epstopdf, epsfig}
\usepackage{amsfonts,amsmath,amssymb}
\usepackage{bm}
\newcommand\pages[2]{#1}
\newcommand{\mockalph}[1]{}

\shorttitle{The magnetic shear-current effect}

\title{The magnetic shear-current effect: generation of large-scale magnetic fields by the small-scale dynamo}

\author{J. Squire\aff{1,}\aff{3}\corresp{\email{jsquire@caltech.edu}} \and A. Bhattacharjee\aff{2,}\aff{3}}

\affiliation{\aff{1}TAPIR, Mailcode 350-17, California Institute of Technology, Pasadena, CA 91125, USA
\aff{2}Department of Astrophysical Sciences and Princeton Plasma Physics Laboratory, Princeton University, Princeton, NJ 08543
\aff{3}Max Planck/Princeton Center for Plasma Physics, Department of Astrophysical Sciences and Princeton Plasma Physics Laboratory, Princeton University, Princeton, NJ 08543, USA}

\begin{document}

\maketitle

\begin{abstract}
A novel large-scale dynamo mechanism, the \emph{magnetic shear-current effect}, is discussed and explored. The effect relies on the interaction of magnetic fluctuations with a mean shear flow, meaning the 
saturated state of the small-scale dynamo can drive a large-scale dynamo---in some sense 
the inverse of dynamo quenching.
 The dynamo is nonhelical, with the mean-field $\alpha$ coefficient zero, and is caused
by the interaction between an off-diagonal component of the turbulent resistivity and the stretching of the large-scale field by shear flow.
Following up on previous numerical and analytic work,  this paper presents further details 
of the numerical evidence for the effect, as well as an heuristic description of
how magnetic fluctuations can interact with shear flow to produce the required electromotive force. 
The pressure response of the fluid is fundamental to this mechanism, which helps explain
why the magnetic effect is stronger than its kinematic cousin, and the basic idea 
is related to the well-known lack of turbulent resistivity quenching by magnetic
fluctuations.  As well as being interesting for its applications to general high Reynolds 
number astrophysical turbulence, where
strong small-scale magnetic fluctuations are expected to be prevalent, the magnetic
shear-current effect is a likely candidate for large-scale dynamo in the 
unstratified regions of ionized accretion disks. Evidence for this is discussed, as well as 
future research directions and the challenges involved with  understanding details of the effect in astrophysically 
relevant regimes. 
\end{abstract}

\section{Introduction}
Magnetic fields pervade the universe. From the scales of planets up
to galaxy clusters and beyond, they are not only ubiquitous but have
also proven dramatically important in a wide variety of
astrophysical and geophysical processes. Despite this, our understanding
the mechanisms that lead to their creation and sustenance is hazy, and improving 
this remains an outstanding theoretical challenge. 
Much of the theory of field generation focuses on \emph{turbulent dynamo}, in which 
magnetic fields are stretched and twisted by turbulent fluctuations in such a way as to increase their strength, resulting in exponential instability. Through this process, very 
small seed fields---arising, for example, from the Biermann battery or 
kinetic instabilities---might be amplified enormously by plasma motions to the 
levels seen throughout the universe today. 

 Interestingly, magnetic fields are generically
observed to be correlated over larger scales than the underlying fluid
motions, and such \emph{large-scale dynamos} are of vital importance for explaining astrophysical fields. 
The classic mechanism
to allow such behavior is the \emph{kinematic} $\alpha$ \emph{effect}\footnote{The term kinematic denotes
the situation where velocity fluctuations are unaffected by the magnetic field.} \citep{Moffatt:1978tc,Krause:1980vr}. Here,
the small-scale fluid turbulence interacts with a  large-scale magnetic field in such a way 
that an electromotive force (EMF, represented by $\bm{\mathcal{E}}$) is created in proportion the magnetic field itself ($\bm{\mathcal{E}}\sim\alpha \bm{B}$),
potentially causing an instability to develop. To allow such behavior,
the turbulence must break statistical symmetry in some way, either through a net helicity or 
 through stratification. However,  various problems with large-scale $\alpha$ dynamos
become apparent when one considers how field growth 
 rates change with the scale of the field---specifically,  the smallest scales always grow the most rapidly 
 \citep{Kulsrud:1992ej,Boldyrev:2005ix}.
 In addition, as a consequence of the conservation of magnetic
 helicity, these small-scale magnetic fields act to decrease the large-scale growth 
 rate in a way that scales very unfavorably to high Reynolds numbers---the problem of ``catastrophic quenching'' \citep{Gruzinov:1994ea,Bhattacharjee:1995ip,Cattaneo:2009cx}. While a 
 variety of solutions to such problems have been explored, primarily focused on 
 the transport of magnetic helicity \citep{Blackman:2002fe,Vishniac:2001wo,Subramanian:2004gm,Ebrahimi:2014jt,Tobias:2014ek}, the scaling of $\alpha$ dynamos to astrophysically relevant regimes is
 still far from understood. Such issues are not necessarily confined to the $\alpha$ effect either.  Above
 even moderate Reynolds numbers, the fast-growing small-scale dynamo (field generation on scales 
 at and below that of the fluid turbulence; \citealt{Schekochihin:2007fy}) implies that 
 velocity fluctuations should \emph{always} be accompanied by magnetic fluctuations of a similar
 magnitude \citep{Schekochihin:2004gj}. This challenges the 
 relevance of the classical kinematic dynamo picture \citep{Cattaneo:2009cx}, which focuses purely 
 on the properties of the small-scale velocity fields.
 
 In this paper---as well as in \citet{HighRm} (hereafter \defcitealias{HighRm}{Paper I}Paper I), \citet{LowRm} (hereafter \defcitealias{LowRm}{Paper II}Paper II), and 
 \citet{Analytic} (hereafter \defcitealias{Analytic}{Paper III}Paper III)---we suggest and explore 
 a new dynamo mechanism in which the small-scale \emph{magnetic fluctuations}, in combination
 with a background shear flow, are 
 the primary driver of the large-scale field growth. Termed the ``magnetic shear-current effect,'' by
 analogy to earlier kinematic suggestions \citep{Urpin:1999wl,Rogachevskii:2003gg}, the effect is
 nonhelical (the dynamo $\alpha$ coefficient is zero),  and is driven by an off-diagonal component of the turbulent resistivity tensor. 
There are two principal reasons for our interest in this effect. The first is that the mechanism is not an $\alpha$ effect, which 
implies that the dynamo can operate in turbulence with a high degree of symmetry. This makes
it a possible mechanism to explain the dynamo seen in the central regions of accretion
disk simulations \citep{Hawley:1996gh,Brandenburg:1995dc}, and we have seen  good evidence
that this is indeed the case (see Sec.~\ref{sec:discussion}, as well as \citealt{Squire:2015fk}). The second reason for our interest in the magnetic
shear-current effect stems from the intriguing possibility of a large-scale 
dynamo being \emph{driven} by the saturated state of the small-scale dynamo. In some sense, this 
is the inverse of the quenching described in the previous paragraph---the small-scale dynamo, far from 
quenching large-scale growth, is its primary driver. 
Such a large-scale dynamo paradigm is far removed from classical kinematic theory, relying on
saturation of the small-scale turbulent fields. Accordingly, the magnetic shear-current effect
is an inherently nonlinear dynamo mechanism \citep{Tobias:2011da}, although it can be driven  
by a turbulent velocity field rather than resulting from the nonlinear development of a laminar instability.

Proving the existence and importance of a dynamo instability is tricky: numerical 
simulations of turbulence are necessarily noisy, one is limited in available Reynolds numbers (and
thus the ability to prove a dynamo will remain active at high values), and 
 when large-scale field growth is observed it can be difficult to show convincingly that 
 it is not some other (possibly unknown) mechanism that is responsible. 
These problems are exacerbated in the magnetically driven case studied in this work. In particular, due to 
the finite size of any numerically realizable mean-field average, the large-scale field will 
 quickly come into equipartition with the turbulent bath of fluctuations, robbing 
the researcher of the ability to study the dynamo during a long period of exponential growth.
In other words, the dynamo will very quickly transition into its saturated state (where the
large-scale fields have a strong influence on the small-scale turbulence), exacerbating  
measurement of the properties of the linear growth phase, or even the observation of its qualitative behavior.
For these reasons, we have attempted to tackle the problem from a variety of different angles,
including analytically with the second-order correlation approximation \citepalias{Analytic}, through quasi-linear theory and statistical simulation  \citepalias{LowRm}, and using 
direct numerical simulations  \citepalias{HighRm,LowRm}.  We also employ the novel technique
of using an \emph{ensemble}  of simulations to study the statistics of 
the mean field without taking  time averages. Our hope is that with this variety of methods,
which all lead to the same general conclusions, we present convincing evidence for the existence of 
the magnetic shear-current effect and its potential importance in astrophysical dynamo theory.

The present paper serves two purposes. The first is to give a more heuristic and physical 
description of the magnetic shear-current effect, which is done throughout Sec.~\ref{sec:The physical mechanism}. 
Following a basic description of the mechanism in the language of mean-field dynamo theory, we describe (with diagrams and simple explanations) how magnetic fluctuations, interacting with a large-scale magnetic field and shear flow, can generate the correlated velocity perturbations that are required for a mean-field dynamo instability. 
Interestingly, we find that the pressure response of the velocity fluctuations is fundamental to the 
operation of the dynamo, and simple arguments based on the directions of induced perturbations explain
qualitatively why one might expect the magnetic effect to be stronger than the kinematic effect.
The second purpose of this paper, discussed in Sec.~\ref{sec:numerical evidence}, is to expand upon, and provide further details for, the analysis and simulations
 presented in \citetalias{HighRm}. In particular, these simulations demonstrate for the first time (so far as we are aware) that the saturated state
 of the small-scale dynamo can \emph{drive} a large-scale dynamo. Our method for showing this involves measuring the
transport coefficients before and after small-scale dynamo saturation. This illustrates that strong magnetic fluctuations
can decrease, and change the sign of, a particularly important component of the tensorial turbulent resistivity (termed $\eta_{yx}$  throughout the text), in a way that is 
consistent with observed mean-field evolution. Since the methods used to show 
this are somewhat nonstandard, considerable effort is put into explaining these and 
ensuring that the coefficients are determined accurately. 
This is done both through direct comparison with standard methods in lower-Reynolds-number kinematic dynamos (appendix~\ref{app:verification}), and by using the measured coefficients to solve for the expected large-scale field evolution. 

Finally, in Sec.~\ref{sec:discussion}, we conclude and present a more in-depth discussion of why the 
magnetic shear-current effect is  interesting as a mechanism for large-scale dynamo. This includes  some analysis of the evidence for the effect's  importance in driving the dynamo in the central regions of
accretion disks, which is primarily based on the Prandtl number dependence of its nonlinear saturation \citep{Squire:2015fk}.

\section{The physical mechanism for the magnetic shear-current effect}\label{sec:The physical mechanism}
In this section we describe how homogeneous nonhelical magnetic fluctuations, 
influenced by a large-scale shear flow and magnetic 
field gradient, can generate an EMF that 
acts to reinforce the large-scale magnetic field.  We shall start by describing
the form of the EMF that allows for such behavior, as well as constraints due to the symmetries of
the system, then consider a simplified cartoon picture for how the interaction of 
magnetic fluctuations with velocity shear and a large-scale field gradient might 
produce this EMF.

All studies in this work are carried out in the context of the incompressible MHD 
equations with a background shear flow $\bm{U}_{0}=-Sx\hat{\bm{y}}$,
\begin{subequations}
\begin{align}
\frac{\partial\bm{U}_{T}}{\partial t}  &-Sx\frac{\partial\bm{U}_{T}}{\partial y}+\left(\bm{U}_{T}\cdot\nabla\right)\bm{U}_{T}+2\Omega\bm{\hat{z}}\times\bm{U}_{T}+\nabla p\nonumber \\
&\qquad \qquad \quad \!=   SU_{Tx}\bm{\hat{y}}+\bm{B}_{T}\cdot\nabla\bm{B}_{T}+\bar{\nu}\nabla^{2}\bm{U}_{T}+\bm{\sigma}_{\bm{u}}, \\
\frac{\partial\bm{B}_{T}}{\partial t} & -Sx\frac{\partial\bm{B}_{T}}{\partial y}=-SB_{Tx}\bm{\hat{y}}+\nabla\times\left(\bm{U}_{T}\times\bm{B}_{T}\right)  +\bar{\eta}\nabla^{2}\bm{B}_{T},\label{eq:induction} \\
 & \nabla\cdot\bm{U}_{T}=0,\;\;\;\nabla\cdot\bm{B}_{T}=0.
\end{align}\label{eq:MHD}\end{subequations}Here $\Omega$ is a mean rotation of the frame, 
and $\bar{\nu}$ and
$\bar{\eta}$ are the normalized viscosity and resistivity respectively.
Since all quantities are normalized to one it is convenient to define
$\mbox{Re}=1/\bar{\nu}$ and $\mbox{Rm}=1/\bar{\eta}$ for the Reynolds
and magnetic Reynolds number, and their ratio is the Prandtl number $\mathrm{Pm}=\mathrm{Rm}/\mathrm{Re}$.  $\bm{\sigma}_{\bm{u}}$ denotes a 
nonhelical driving noise source, white in time, which
can be used to generate an homogenous bath of small-scale velocity fluctuations. 
$\bm{U}_{T}$ and $\bm{B}_{T}$ in Eq.~\eqref{eq:MHD} are simply the
standard turbulent velocity and magnetic fields ($\bm{U}_{T}$
is the velocity not including the background shear).
Throughout this work
we consider initially homogenous turbulence with zero average helicity.

\subsection{Nonhelical dynamo mechanisms}\label{sub:Nonhelical dynamo mechanisms}

To examine field generation mechanisms in this geometry, 
it is helpful to start by defining mean and fluctuating
fields through the relation $\bm{B}_{T}=\overline{\bm{B}}_{T}+\bm{b}=\bm{B}+\bm{b}$.
Here $\bar{\cdot}$ is the \emph{mean-field average}, which is taken to be a spatial average over $x$ and
$y$.  An average of the induction equation (Eq.~\eqref{eq:induction}) leads 
leads to the well-known mean-field dynamo equations for the mean magnetic
field $\bm{B}$ \citep{Moffatt:1978tc,Krause:1980vr},
\begin{equation}
\partial_{t}\bm{B}=\nabla\times\left(\bm{U}_{0}\times\bm{B}\right)+\nabla\times\bm{\mathcal{E}}+\frac{1}{\mathrm{Rm}}\nabla^{2}\bm{B}.\label{eq:genMF}
\end{equation}
Here $\bm{\mathcal{E}}=\overline{\bm{u}\times\bm{b}}$ is the EMF, which provides the connection between the small-scale turbulence and large-scale fields. If we assume scale separation between 
the mean and fluctuating fields, a Taylor expansion of $\bm{\mathcal{E}}$ leads
to the form \begin{equation}
\mathcal{E}_{i}=\alpha_{ij}B_{j}-\eta_{ij}J_{j}+\cdots,\label{eq:E expansion}\end{equation}
where $\alpha_{ij}$ and $\eta_{ij }$ are the transport coefficients, and 
the lack of $(x,y)$ dependence of the mean fields  has been used to reduce the number of $\eta$ coefficients from $27$ to $4$ (note that $B_{z}=0$). In the case where the mean fields can be 
considered a small perturbation to some background turbulent state
specified by statistics of $\bm{u}$ and $\bm{b}$ (which are influenced by shear and rotation), 
$\alpha$ and $\eta$ must be independent of $\bm{B}$. 

Due to reflectional symmetry, with a nonhelical forcing function $\bm{\sigma}_{\bm{u}}$, the
$\alpha_{ij}$ coefficients are constrained to vanish on average in this geometry. Instead, 
we shall study the possibility of a mean-field dynamo that arises purely from the off-diagonal components of
$\eta_{ij}$, which can be nonzero due to the anisotropy of the turbulence. 
Combining Eqs.~\eqref{eq:genMF} and \eqref{eq:E expansion}, one obtains 
\begin{subequations}
\begin{gather}
\partial_{t}B_{x}=-\alpha_{yx}\partial_{z}B_{x}-\alpha_{yy}\partial_{z}B_{y}-\eta_{yx}\partial_{z}^{2}B_{y}+(\eta_{yy}+\bar{\eta})\partial_{z}^{2}B_{x}, \\
\partial_{t}B_{y}=-SB_{x}+\alpha_{xx}\partial_{z}B_{x}+\alpha_{xy}\partial_{z}B_{y}-\eta_{xy}\partial_{z}^{2}B_{x}+(\eta_{xx}+\bar{\eta})\partial_{z}^{2}B_{y}
\end{gather}\label{c4:eq:SC sa eqs}\end{subequations}where the time
average of the $\alpha_{ij}$ components must vanish.
From
these equations (with $\alpha_{ij}=0$), it is straightforward to show that an eigenmode with the spatial structure
$B_{i}=B_{i0}e^{ikz}$ has the growth rate 
\begin{equation}
\gamma_{\eta}=k\sqrt{\eta_{yx}\left(-S+k^{2}\eta_{xy}\right)}-k^{2}\eta_{t},\label{c4:eq:gamma SC}
\end{equation}
where we have set $\eta_{yy}=\eta_{xx}=\eta_{t}$ for simplicity.
Neglecting $\eta_{xy}$ by assuming $\left|k^{2}\eta_{xy}\right|\ll S$
(for all $k$ for which scale separation  holds), one finds that positive
dynamo growth is possible if $-S\eta_{yx}>0$ and $k\sqrt{-\eta_{yx}S}>k^{2}\eta_{t}$.
The physical mechanism for the instability involves the $B_{x}$  generated 
by $B_{y}$ (through $-\eta_{yx}\partial_{z}^{2}B_{y}$) feeding back on $B_{y}$ through 
stretching by the mean shear flow (the $-SB_{x}$ term in Eq.~\eqref{c4:eq:SC sa eqs}).
Thus the possibility of such a nonhelical dynamo rests crucially on the phase 
between $B_{x}$ and $B_{y}$ and therefore on
the transport coefficient $\eta_{yx}$, which must be less than zero. 


Whether $\eta_{yx}$ is positive or negative depends on the properties 
of the turbulence in question, in particular on the sign of $(\overline{\bm{u}\times \bm{b}})_{y}$
that arises in the presence of a ${B}_{y}$ gradient. 
The standard kinematic approach in dynamo theory has been to consider 
strong underlying hydrodynamic fluctuations (denoted by $\bm{u}_{0}$), which
generate $\bm{b}$ fluctuations through interaction with $\nabla \bm{B}$ (and $\bm{B}$).
Although various early analytic works argued for a kinematic shear-current dynamo of this type \citep{Urpin:1999wl,Urpin:2002ct,Rogachevskii:2003gg}, 
subsequently several authors found that kinematically $\eta_{yx}>0$ (at least at low $\mathrm{Rm}$) and 
thus concluded that a coherent kinematic dynamo  
cannot explain the field generation observed in numerical experiments (\citealt{Radler:2006hy,Brandenburg:2008bc,Singh:2015kt}; \citetalias{LowRm}).
Here we argue instead that strong homogenous magnetic fluctuations (denoted by $\bm{b}_{0}$) can
generate $\bm{u}$ fluctuations with the required correlations to cause a negative
$\eta_{yx}$. Such $\bm{b}_{0}$ fluctuations should be ubiquitous in MHD turbulence at high Reynolds numbers, 
since the small-scale dynamo will be unstable (with a large growth rate set 
by the smallest scales in the turbulence), creating a turbulent state with $\bm{b}_{0}\sim\bm{u}_{0}$ \citep{Schekochihin:2004gj}.

Before continuing, it is worth mentioning another possibility for large-scale field generation in this
geometry---the so-called, stochastic-$\alpha$ effect. This arises through {fluctuations} in the $\alpha_{ij}$ coefficients, even though their mean must vanish 
\citep{Vishniac:1997jo,Heinemann:2011gt,Mitra:2012iv}.
This dynamo is not mean-field in the usual sense since it relies on
the finite size of the system to cause the $\alpha$ fluctuations that lead to mean-field growth; nonetheless,
given that the universe is sampling a single realization of turbulence,
\emph{not} the ensemble average, such effects could be entirely physical.
(That said, one consequence of this incoherent dynamo mechanism is that the growth rate
can be arbitrarily increased or decreased by changing the volume of the mean-field 
average, which
hints that coherent  effects should dominate when a very large range of scales are present.) 
While we shall not examine the stochastic-$\alpha$ effect in detail in this work (see \citetalias{LowRm}), 
it is important to be mindful of the possibility, since it complicates the analysis of simulation results where
large-scale field growth is observed. 
One distinguishing feature from the shear-current effect is that 
$\bm{B}\left(z,t\right)$ cannot have a constant phase in
time as it grows, since the average of $\bm{B}$ over an ensemble of realizations vanishes,
implying $\bm{B}$ must be uncorrelated with itself after $t\gtrsim(k^{2}\eta_{t})^{-1}$\footnote{This condition may be altered if one considers the effects of magnetic helicity conservation, which may cause a local magnetic $\alpha$ effect as the large-scale field grows \citep{Brandenburg:2008bc}. However, this also causes coupling between different mean-field modes, and the detailed consequences of such an effect remain unclear.}. 
More information, including analyses of the relative importance of the coherent and
incoherent shear dynamo mechanisms in low-$\mathrm{Rm}$ systems, can be found in \citetalias{LowRm}.

\subsection{The mechanism for the magnetic shear-current effect}\label{sec:MSC}

In this section we discuss the mean field generation mechanism of the magnetic
shear-current effect. The stability analysis given in Sec.~\ref{sub:Nonhelical dynamo mechanisms} makes it clear that we require $\eta_{yx}<0$ for a coherent dynamo instability. In the present context, with both the mean magnetic field and flow in the  $y$ direction and their prescribed spatial dependencies (see Fig.~\ref{fig:SCResis Diagram}, left panel), this is equivalent to requiring that the $y$ component of the turbulent EMF be negative. The challenge is then for us to explain how this can come about in the present geometry.
The cartoon picture that we present has its origins in the analytic ``second-order
correlation approximation'' (SOCA) calculations presented in \citetalias{Analytic}. In particular, by 
selectively removing terms from the calculation and examining the effects on
the final $\eta_{yx}$, one can unambiguously determine from where the effect arises
(at least within the quasi-linear approximation). Most importantly, this exercise shows that
the magnetic shear-current effect arises exclusively from the \emph{pressure response}
of the velocity fluctuations.  The mechanism is fundamentally related to 
the lack of turbulent resistivity quenching  by the magnetic field (often referred to as a
lack of ``$\beta$ quenching''; see \citealt{Gruzinov:1994ea} and \citealt{Bhattacharjee:1995ip}), which
results from a cancellation between a turbulent magnetic resistivity (of the
same form as kinematic turbulent resistivity), and an equal and opposite contribution 
from the pressure response \citep{Avinash:1991fu}. 

We divide our discussion up as answers to three questions: (1) How do we generate the fluctuations needed to support the required EMF? (2) What happens in the absence of flow shear?  and (3) What happens in the presence of flow shear?

\subsubsection{How do we generate the fluctuations needed to support the required EMF?}\label{sec:generating fluctuations}

The fluctuations needed to support our physical picture are magnetically driven. 
In contrast to kinematic dynamos, the Maxwell stress $\bm{B}_{T}\cdot \nabla \bm{B}_{T}$ is fundamental 
for a magnetically driven dynamo, since this is required to generate $\bm{u}$ from $\bm{b}$ (in 
the same way the Lorentz force $\nabla \times (\bm{U}_{T}\times \bm{B}_{T})$ generates correlated
$\bm{b}$ fluctuations in kinematic dynamos). 
Such dynamos can still be analyzed linearly if one assumes that the interaction of fluctuations 
with mean fields is more important for the EMF than the interaction with themselves; that is,
\begin{equation}
\bm{b}\cdot \nabla \bm{B}+\bm{B}\cdot \nabla \bm{b} \quad\text{is more important than} \quad\bm{b}\cdot \nabla \bm{b} - \overline{\bm{b}\cdot \nabla \bm{b}}.\label{eq:QL approx}
\end{equation}
This approximation---which along with a similar approximation for the Lorentz force, is the basis for SOCA---is  valid only at low Reynolds numbers and nonzero mean fields, but allows 
one to consider how small-scale eddies and field loops would interact with
large-scale field and flow gradients in a relatively straightforward way.
Note that, ``is more important'' in Eq.~\eqref{eq:QL approx}  refers to the terms' relative importance 
for the generation of an EMF that is correlated with $\bm{B}$ (this correlation is necessary for a large-scale dynamo).
Since only the part of $\bm{b}$ that is influenced by $\bm{B}$ can contribute 
to this correlation,  it seems reasonable to surmise that results should be qualitatively applicable
outside their true validity range. In other words, since the interaction of $\bm{b}$ with $\bm{B}$
is the cause of the magnetic shear-current effect in the first place, we shall focus on this (rather than the
much more complicated nonlinear terms)  for the development 
of our simple cartoon model.
\begin{figure}
\begin{center}
\includegraphics[width=0.5\columnwidth]{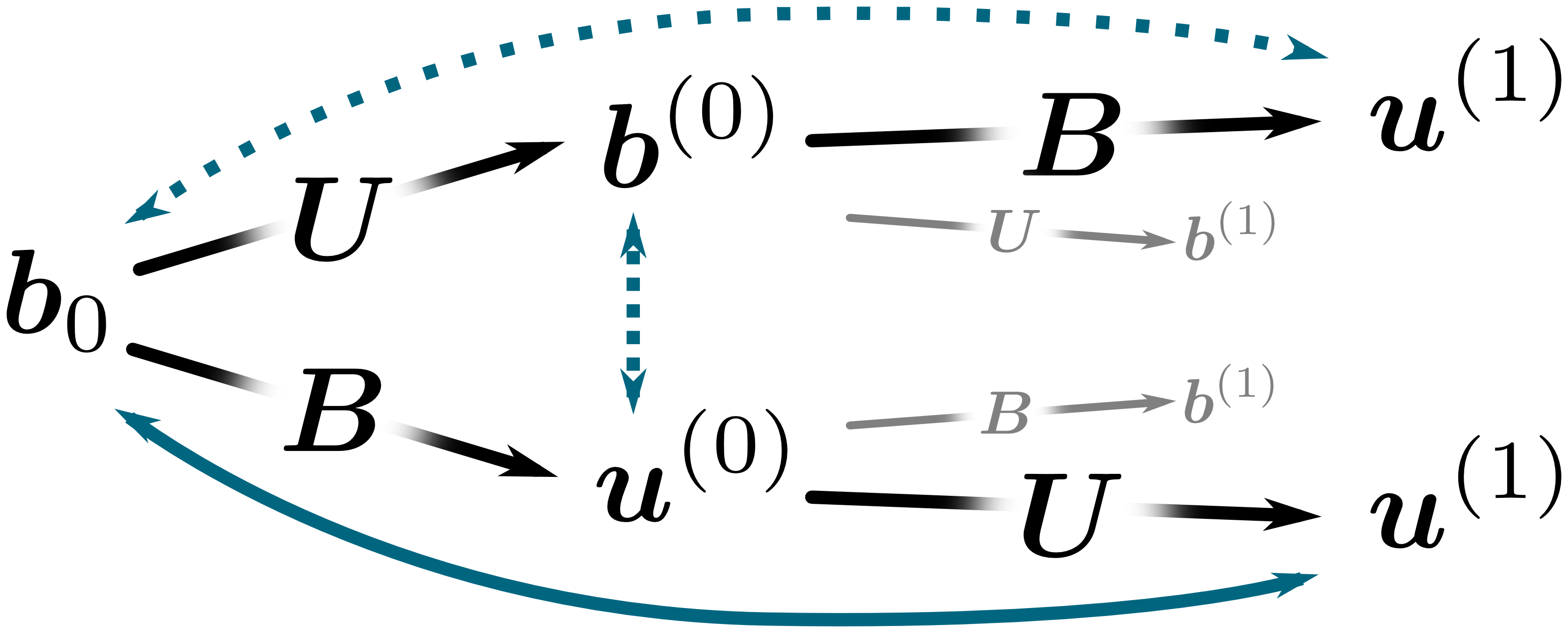}
\caption{Depiction of the interactions between
fluctuations ($\bm{b}_{0}$, $\bm{u}^{(i)}$, and $\bm{b}^{(i)}$) and  a mean magnetic field $\bm{B}$ or a shear flow $\bm{U}$, that can lead to a nonzero shear-current effect through $\overline{\bm{u}\times \bm{b}}$, starting from strong 
homogenous magnetic fluctuations $\bm{b}_{0}$. Here the straight black arrows, with either $\bm{B}$ or $\bm{U}$, 
depict an interaction that creates one fluctuating field from another, which will be correlated with the original fluctuation and 
thus can contribute to the EMF (for instance $\bm{u}^{(0)}\sim \tau_{c }\bm{B}\cdot \nabla \bm{b}_{0}+ \tau_{c} \bm{b}_{0}\cdot \nabla \bm{B}$). The double headed (blue) arrows indicate the lowest order
combinations of $\bm{b}_{0}$, $\bm{u}^{(i)}$, and $\bm{b}^{(i)}$ that can lead to nonzero $\eta_{yx}$, with
the interaction studied in Sec.~\ref{sec:MSC} shown by the solid line. \label{fig:Interactions} }
\end{center}
\end{figure}

The shear-current effect requires both a field gradient 
and a flow gradient (shear flow). Thus, any perturbation 
$\bm{b}_{0}$ (arising as part of the bath of statistically homogenous magnetic fluctuations) must 
interact with both $\bm{U}$ and $\bm{B}$ to generate a $\bm{u}$
fluctuation. The possible ways in which this can happen are illustrated in Fig.~\ref{fig:Interactions}, where
the notation is the same as that used in \citetalias{Analytic}, with $\bm{f}^{(0)}$ indicating a field 
that arises directly from the interaction of $\bm{b}_{0}$ (or $\bm{u}_{0}$) with the mean fields, 
and $\bm{f}^{(1)}$ indicating one that arises through $\bm{f}^{(0)}$.
In addition, we use $(\cdot )_{b}$ to denote the part of a transport coefficient that is due to homogenous magnetic fluctuations; for example, $(\eta_{yx})_{b}$.
From the momentum equation, 
a $\bm{b}$ perturbation can generate a $\bm{u}$ perturbation through $\bm{b}\cdot \nabla \bm{B}+\bm{B}\cdot \nabla \bm{b}$, while 
a $\bm{u}$ perturbation can generate a $\bm{u}$ perturbation through $-\bm{u}\cdot \nabla \bm{U}-\bm{U}\cdot \nabla \bm{u}$. 
Similarly, from the induction equation a $\bm{b}$ perturbation is generated through 
either a $\bm{u}$ perturbation ($\bm{B}\cdot \nabla \bm{u}$), or through a 
$\bm{b}$ perturbation ($-\bm{U}\cdot \nabla \bm{b}$). We see from Fig.~\ref{fig:Interactions}
that there are three possibilities for contributing to $(\eta_{yx})_{b}$: 
$\bm{u}^{(0)}\times \bm{b}^{(0)}$,  $(\bm{u}^{(1)}\times \bm{b}_{0})_{1}$, 
and $(\bm{u}^{(1)}\times \bm{b}_{0})_{2}$.
Here $(\bm{u}^{(1)}\times \bm{b}_{0})_{1}$ refers to the pathway for 
generating $\bm{u}^{(1)}$ through $\bm{u}^{(0)}$ (shown by the solid arrow in Fig.~\ref{fig:Interactions}), 
while $(\bm{u}^{(1)}\times \bm{b}_{0})_{2}$ refers to the pathway 
through $\bm{b}^{(0)}$ (shown by the top dashed arrow).
  Out of these, we have determined 
from the calculations in \citetalias{Analytic} that $(\bm{u}^{(1)}\times \bm{b}_{0})_{1}$ is both the simplest and contributes the most to 
$(\eta_{yx})_{b}$. In particular, the mechanism does not directly rely on dissipation to generate the
required correlations, as will be seen below\footnote{To be more specific, in the SOCA calculations, 
this contribution does not involve $k$ derivatives of the functions $E_{\eta}=1/(i\omega -\bar{\eta}k^{2})$
or $N_{\nu}=1/(i\omega -\bar{\nu}k^{2})$ (which are zero at $\bar{\nu}=0$ or $\bar{\eta}=0$), while the other two contributions do.}. We have found empirically that the $\bm{u}^{(0)}\times \bm{b}^{(0)}$ contribution is moderate in size (generally a factor of $\sim 2$ smaller than $\bm{u}^{(1)}\times \bm{b}_{0}$) 
and also always negative, while the $(\bm{u}^{(1)}\times \bm{b}_{0})_{2}$ contribution (dotted line in Fig.~\ref{fig:Interactions}) can change sign but is much smaller in magnitude.

\subsubsection{What happens in the absence of flow shear?}\label{sub:diagonal resistivity}
\begin{figure}
\begin{center}
\includegraphics[width=1.0\columnwidth]{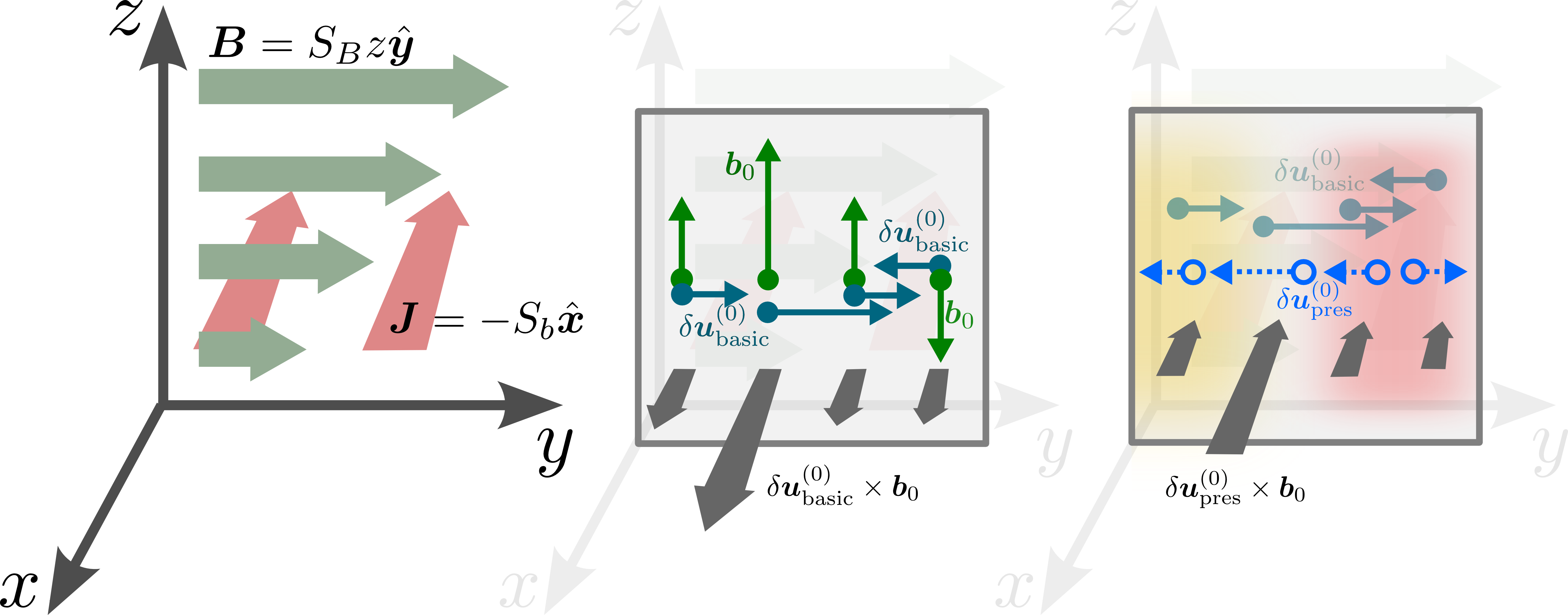}
\caption{Graphical illustration of the mean-field resistivity---or lack thereof---generated
by homogenous small-scale magnetic fluctuations, with the geometry of the mean field illustrated in the left-hand panel. The middle panel shows how $b_{0z}$ perturbations (from an homogeneous turbulent bath) lead to a $\bm{u}$ perturbation (labelled $\delta \bm{u}^{(0)}_{\mathrm{basic}}$)
through $\bm{b}\cdot \nabla \bm{B} =S_{B}b_{0z}\hat{\bm{y}} $, resulting in an EMF in the $-\bm{J}$ direction. The 
right-hand panel shows how the pressure response to this $\delta \bm{u}^{(0)}_{\mathrm{basic}}$ (labelled $\delta \bm{u}^{(0)}_{\mathrm{pres}}$), which arises due to its 
nonzero divergence (yellow and red shaded regions for $\nabla \cdot \delta \bm{u}^{(0)}_{\mathrm{basic}}>0$ and $\nabla \cdot \delta \bm{u}^{(0)}_{\mathrm{basic}}<0$ respectively),
leads to an EMF that opposes that from $\delta \bm{u}^{(0)}_{\mathrm{basic}}$. A more careful calculation shows that the cancellation is exact (in incompressible turbulence at low $\mathrm{Rm}$), so the turbulent resistivity  due to magnetic fluctuations vanishes. See text for further discussion. 
 \label{fig:DiagResis Diagram}}
\end{center}
\end{figure}

As mentioned above, in the absence of flow shear, there is no quenching of the turbulent resistivity.
This effect---which could also be stated as $(\eta_{xx})_{b}=(\eta_{yy})_{b}=0$ in the notation of Eq.~\eqref{c4:eq:SC sa eqs}---arises through the pressure response of the fluid.
We feel
it helpful to first explain this mechanism in more detail, since the form of the pressure response 
has not been discussed in detail in previous literature  (so far as we are aware)\footnote{
\cite{Yokoi:2013di} gives a slightly different model for the negative contribution to the magnetic 
resistivity, based on small-scale current perturbations creating a magnetic pressure. This model is  
similar to that presented here but does not directly include the fluid pressure (it includes the magnetic pressure), which we have seen
to be important by studying the relevant terms in SOCA calculations.}
and the magnetic shear-current effect is essentially an extension of this. 
As can be seen using SOCA (or the $\tau$~approximation; see \citealt{Radler:2003gg}), the effect
occurs because
the pressure response has an equal and opposite effect to the primary velocity perturbation \citep{Avinash:1991fu}. 
This behavior is illustrated graphically in Fig.~\ref{fig:DiagResis Diagram}, which shows the response 
of the fluid to a magnetic perturbation in the linearly varying magnetic field $\bm{B}=S_{B}z \hat{\bm{y}}$.
Due to the mean-field geometry, the velocity perturbation $\delta \bm{u}^{(0)}_{\mathrm{basic}} \sim \tau_{c} \bm{b} \cdot \nabla \bm{B}$ (where $\tau_{c}$ is some turbulent correlation time) is simply $S_{B} b_{0z} \hat{\bm{y}}$; i.e., only the $z$ component of $\bm{b}_{0}$ contributes. 
Note that the other contribution  $\delta \bm{u}^{(0)}_{\mathrm{basic}} 
\sim \tau_{c} \bm{B}\cdot \nabla \bm{b}$, will only contribute directly to the EMF if there 
is a mean correlation between $\bm{b}$ and $\nabla \bm{b}$, which occurs 
if there is net current helicity (this term is the origin of the magnetic $\alpha$ effect)\footnote{The
general situation is a little more complex than this. Fourier transformed, a term of the 
form $\bm{B }\cdot \nabla \bm{b}$ becomes $i \bm{B}\cdot \bm{k}\, \bm{b}-(\nabla \bm{B})_{jl}k_{j}\partial_{k_{l}}b_{i}+\cdots$ (where the Einstein summation convention is used). 
Without helicity the first term does not contribute when averaged over a domain,
but the second term can in general be nonzero. However, contributions
of this form generally seem to be smaller in magnitude, and its dependence on the $\bm{k}$ derivative of the
fluctuations makes it troublesome to arrive at a simple cartoon picture. See \citetalias{Analytic}
for more detail on the mathematics of the calculation.}.
Obviously, the perurbation $\delta \bm{u}^{(0)}_{\mathrm{basic}} \sim S_{B} b_{0z} \hat{\bm{y}}$ 
is correlated with $\bm{b}_{0}$ and it is straightforward to see (see middle panel of Fig.~\ref{fig:DiagResis Diagram})
that a net $\bm{\mathcal{E}}$ is created in the $\hat{\bm{x}}$ direction, opposite to the mean current and thus 
acting as a turbulent dissipation for the mean field. 

However, as is clear from Fig.~\ref{fig:DiagResis Diagram}, the $\delta \bm{u}^{(0)}_{\mathrm{basic}}$ 
perturbation is not divergence free, given any $y$ variation in $b_{0z}$. In the third panel 
of Fig.~\ref{fig:DiagResis Diagram}, the shaded regions illustrate where the divergence of 
$\delta \bm{u}^{(0)}_{\mathrm{basic}}$ is positive (yellow) or negative (red). Given the
incompressibility of the fluid, a nonzero divergence is not possible, and the $\nabla p$ term 
responds appropriately, creating a flow perturbation from regions of
negative divergence to positive divergence (mathematically, $ \delta \bm{u}^{(0)}_{\mathrm{pres}} = \nabla^{-2}[ -\nabla (\nabla \cdot \delta \bm{u}^{(0)}_{\mathrm{basic}})]$).
As shown in the third panel of Fig.~\ref{fig:DiagResis Diagram} this perturbation 
is anti-correlated with $\delta \bm{u}^{(0)}_{\mathrm{basic}}$ and thus creates
an oppositely directed EMF, in the $+\bm{J}$ direction. 
Further, since
$\nabla \cdot (\bm{b}\cdot \nabla \bm{B})=\nabla \cdot (\bm{B}\cdot \nabla \bm{b})$,
each of these linear contributions to the Maxwell stress add in the same way to the pressure perturbation, and 
a more careful calculation shows that the effect exactly cancels the original EMF on average.
Given its reliance on the pressure response, the effect will  be reduced in  
a compressible flow (presumably becoming negligible for high Mach number flows), 
and one would expect $\bm{b}_{0}$ fluctuations to  increase the turbulent diffusivity in this case (the magnetic shear-current 
effect will also be less effective in a compressible flow).
Finally, it is worth mentioning that  $z$ variation of the initial $\bm{b}_{0}$ perturbation will 
not contribute since this creates a $(\delta \bm{u}^{(0)}_{\mathrm{pres}})_{z}$ (which 
is zero in a cross product with $b_{0z}$), while $x$ variation of $\bm{b}_{0}$ produces a
$(\delta \bm{u}^{(0)}_{\mathrm{pres}})_{x}$ that is out of phase with the original $b_{0z}$ perturbation.

\begin{figure}
\begin{center}
\includegraphics[width=1.0\columnwidth]{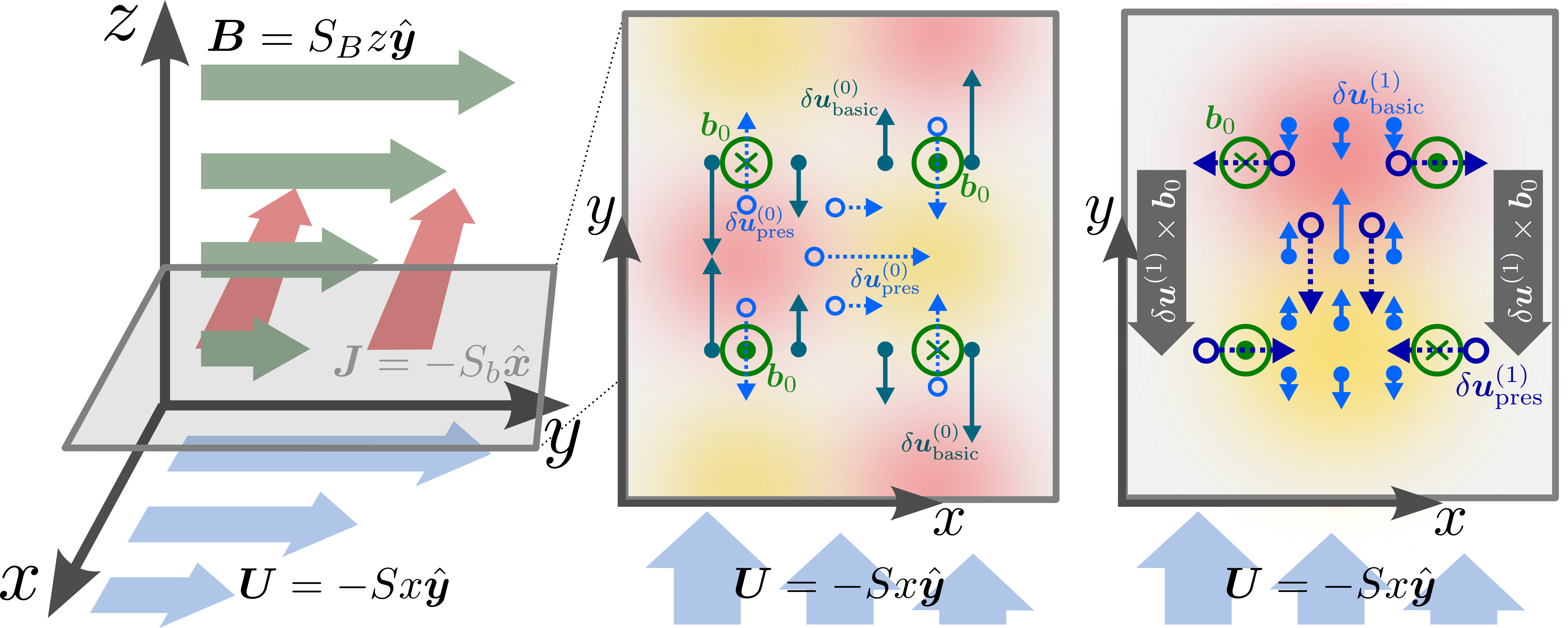}
\caption{Graphical illustration of the magnetic shear-current 
effect, which should be interpreted as follows.  Left-hand panel, the geometry of the mean
field and shear flow.  Middle panel, the  flow perturbation
(both $\delta \bm{u}^{(0)}_{\mathrm{basic}}$ and $\delta \bm{u}^{(0)}_{\mathrm{pres}}$) 
that arises due to $x$, $y$ dependence of the initial $b_{0z}$, before interaction with the shear flow (note 
the rotation of the axes compared to the left panel). Right-hand
panel, the $\delta \bm{u}^{(1)}$ perturbation that arises from $\delta \bm{u}^{(0)}$ due to stretching by the flow, 
which illustrates a correlation between $\delta \bm{u}^{(1)}_{\mathrm{pres}}$ and 
the original $b_{0z}$ structure. The resulting $\bm{\mathcal{E}}$ is pointing
in the $-\hat{\bm{y}}$ direction, corresponding to a negative $\eta_{yx}$. In the middle panel, 
the yellow (red) shading indicates where  $\delta \bm{u}^{(0)}_{\mathrm{basic}}$ has a positive (negative)
divergence, while the shading in the right panel shows the same for $\delta \bm{u}^{(1)}_{\mathrm{basic}}$.  More
information and discussion is given in the main text. \label{fig:SCResis Diagram}}
\end{center}
\end{figure}
\subsubsection{What happens in the presence of flow shear?}
In the presence of flow shear, the cancellation discussed in the previous section leaves a
residual $x$-directed $\bm{u}$ perturbation. This perturbation---which arises from the interaction of the pressure perturbation in Fig.~\ref{fig:DiagResis Diagram} with the 
mean shear, followed by the pressure response to this secondary perturbation---leads to the magnetic shear-current effect. 
This rather complex process is illustrated in graphically in Fig.~\ref{fig:SCResis Diagram}, using 
similar conventions (and color schemes) to Fig.~\ref{fig:DiagResis Diagram}. 
A shear flow in the $\hat{\bm{y}}$ direction is included in addition to the mean field 
$\bm{B}=S_{B}z \hat{\bm{y}}$, which corresponds exactly to the geometry 
discussed in Sec.~\ref{sub:Nonhelical dynamo mechanisms} and Eq.~\ref{c4:eq:SC sa eqs}. 
Recall that $\eta_{yx}<0$ is equivalent to ${\mathcal{E}}_{y}<0$ in this geometry (see Eq.~\eqref{c5:eq:EMF general}).
The second panel in Fig.~\ref{fig:SCResis Diagram} illustrates the same effect 
as shown in Fig.~\ref{fig:DiagResis Diagram}, now including $x$ and $y$ dependence of 
the $b_{0z}$ perturbation. As is evident, even though $\delta \bm{u}^{(0)}_{\mathrm{basic}}$ points
only in the $y$ direction, the pressure response includes equally strong $x$ directed flows, since it 
arises from the spatial dependence of $\nabla \cdot \delta \bm{u}^{(0)}_{\mathrm{basic}}$. 
The resulting $(\delta \bm{u}^{(0)})_{x}$ is out of phase with $\bm{b}_{0}$, so does not contribute 
to an EMF itself, but it is sheared
by the background flow thrdough 
$\delta \bm{u}^{(1)}_{\mathrm{basic}}\sim  -\tau_{c} \bm{u}^{(0)}\cdot \nabla \bm{U}=\tau_{x}S u_{x}^{(0)}\hat{\bm{y}}$, 
which is shown in the third panel of Fig.~\ref{fig:SCResis Diagram}.
Again, since only the $x$ component contributes, $\delta \bm{u}^{(1)}_{\mathrm{basic}}$ is
not divergence free (shaded yellow and red regions for  $\nabla \cdot \delta \bm{u}^{(1)}_{\mathrm{basic}}>0$ and
$\nabla \cdot \delta \bm{u}^{(1)}_{\mathrm{basic}}<0$ respectively). We see that the $x$ component of the pressure response towards (away from)  
regions where $\nabla \cdot \delta \bm{u}^{(1)}_{\mathrm{basic}}>0$ ($\nabla \cdot \delta \bm{u}^{(1)}_{\mathrm{basic}}<0$)
is now correlated and in phase with the original perturbation. Most importantly, its direction 
is such that $\bm{\mathcal{E}}=\delta \bm{u}^{(1)} \times \bm{b}_{0}$ 
is always in the $-\hat{\bm{y}}$ direction, leading to $(\eta_{yx})_{b}<0$. Note that here,
unlike in discussion of Fig.~\ref{fig:DiagResis Diagram}, the effect relies on the $x$ component of the pressure
response (perpendicular to $\delta \bm{u}_{\mathrm{basic}}$), 
which must occur for any perturbation that varies in $x$  because the 
response is the gradient of a scalar field (i.e., $-\nabla p$).

At this point, the reader could be forgiven for viewing the magnetic shear-current 
mechanism explained above with some skepticism---how do we know there
are no opposing mechanisms to cancel out such effects?  The
simplest answer is that we have derived the physical picture in Fig.~\ref{fig:SCResis Diagram}
from the SOCA calculation, by noting that $(\eta_{yx})_{b}$ is unchanged by removal 
of all contributions to the velocity perturbation other than $\nabla p$, and through the exploration
of the different pathways in Fig.~\ref{fig:Interactions}. More physically, 
the reason the pressure is necessary for the shear-current effect arises 
from the mean-field and flow geometry. In particular, if a small-scale fluctuation interacts with either $\bm{U}$ or $\bm{B}$
through 
$\bm{u}\cdot \nabla \bm{U}$, $\bm{u}\cdot \nabla \bm{B}$, $\bm{b}\cdot \nabla \bm{U}$, 
or $\bm{b}\cdot \nabla \bm{B}$,
the resulting perturbation is \emph{always} in the $\pm\hat{\bm{y}}$ direction. Obviously, 
such a perturbation cannot lead to a nonzero ${\mathcal{E}}_{y}$. Thus, $\eta_{yx}$ is
both very important for dynamo action and particularly complicated to generate, because the flow 
and mean field are in the same direction as the required EMF. This explains why $\eta_{yx}$ is seen
to be much smaller than $\eta_{xy}$ in numerical simulation and calculations
(\citealt{Brandenburg:2008bc,SINGH:2011ho}; \citetalias{LowRm}), as well as the capricious nature
of the kinematic shear-current effect (the sign of $(\eta_{yx})_{u}$ may depend on the Reynolds numbers, while
analytic results depend on the closure method used), for 
which these same arguments apply\footnote{There is no
equivalent of the above picture for the kinematic effect. This can be seen 
by the fact that the $\bm{b}$ perturbation that arises from $\bm{b}\cdot \nabla \bm{U}$ is necessarily 
in the $y$ direction (since there is no pressure). This means the $\bm{U}\cdot \nabla \bm{b}$ term,
which involves the derivative of $\bm{b}$ in $\bm{k}$ space,
must be responsible for any $\mathcal{E}_{y}$ (another
possibility is correlations between components of $\bm{u}_{0}$ arising from $\nabla \cdot \bm{u}_{0}=0$). The same is true of the other pathways
in Fig.~\ref{fig:Interactions}.}.
Note that the requirements for $\mathcal{E}_{y}\neq 0$ in Fig.~\ref{fig:SCResis Diagram} 
are very specific---a $y$ variation of the $x$ variation of $b_{0z}$---and
it is straightforward to see that this is the only possibility for generation of a $\delta u_{x}$
in this way.  This implies
we can ignore both the other $\bm{b}$ components and any variation in $z$. 
Thus, although Figs.~\ref{fig:DiagResis Diagram} and \ref{fig:SCResis Diagram}
show the fluid response to a rather specific form for $\bm{b}_{0}$, the important features of the resulting perturbations (shown in the right-hand panels)
are relatively generic for  more general $\bm{b}_{0}$.
Further, since the effect is 
linear, Fourier modes (such as that shown in Fig.~\ref{fig:SCResis Diagram})
can be added to form a more general $\bm{b}$ perturbation,
 and we are  sure to obtain $\mathcal{E}_{y}<0$.  

We have thus seen how magnetic fluctuations can produce a negative $\eta_{yx}$ through their interaction with large-scale field and flow gradients, which can in some cases lead to large-scale dynamo action. 
In addition, the reliance of the effect on the fluid pressure response, as well as the general difficulty of
creating perturbations that create an EMF parallel to both the mean flow and field,  help explain the relative dominance of the 
magnetic  over the kinematic shear-current effect.

\section{Numerical evidence}\label{sec:numerical evidence}
In this section we illustrate numerically that the magnetic shear-current mechanism discussed
above is indeed realizable in MHD turbulence. We show using direct 
numerical simulation that it is possible and realizable to have the small-scale 
dynamo \emph{drive} the growth of the large-scale dynamo.
So far as we are aware, this is the first demonstration of this interesting behavior.

The methods used to illustrate this effect in numerical simulation are somewhat nonstandard in the
dynamo literature. In particular, at each set of physical parameters 
we carry out an ensemble of simulations, each with different noise realizations.
We then measure transport coefficients before and after small-scale saturation in each simulation, which shows (after an ensemble average) that  $\eta_{yx}$ becomes more negative  after 
the saturation of the small-scale dynamo. That this  can drive a coherent 
dynamo is illustrated by qualitative observation of the 
mean-field pattern, as well as solution of the mean-field equations (Eq.~\eqref{c4:eq:SC sa eqs})
using the measured transport coefficients.
The ensemble of simulations is required  as a result of  the relatively short period of large-scale dynamo growth
before nonlinear saturation effects become significant. This is because the large-scale magnetic 
field starts its growth (when the small-scale dynamo saturates) at relatively large amplitudes, 
being in approximate equipartition with the small-scale fluctuations due to 
the finite size of the mean-field average. We shall see that in many cases, the 
growth of the mean field lasts little more than $20\rightarrow 30$ shearing times before 
saturating, and that its behavior can vary substantially between
realizations. Because of this, the ensemble average over simulations is highly advantageous 
for accurate determination of the transport coefficients.  (In other work we have used 
statistical simulation to circumvent this problem,
but this requires a quasi-linear approximation, which eliminates in small-scale 
dynamo; see \citetalias{LowRm})

The method for measuring the transport coefficients from simulation data 
after small-scale dynamo saturation  (termed the ``projection method'')
is also nonstandard, and will be explained in some detail. 
Because test-field methods 
that explicitly include the magnetic fluctuations are rather complex 
and in the early stages of development \citep{Rheinhardt:2010do}, 
we  instead choose to measure 
transport coefficients directly from mean-field and EMF data taken from 
simulations. The method, which is a modified version of that proposed in \citet{Brandenburg:2002cia} and is 
also used with some success in \citet{Racine:2011gh} and \citet{Simard:2013jf},
involves approximately solving $\mathcal{E}_{i}=\alpha_{ij}B_{j}-\eta_{ij}J_{j}$ at each
time-step and taking spatiotemporal averages to obtain transport coefficients. 
To ensure that correct results are obtained, the projection method is checked in two independent ways: First, it is used
to compute transport coefficients for low-$\mathrm{Rm}$ kinematic shear dynamos and 
compared directly to the test-field method (this is presented in appendix~\ref{app:verification}). Second, we solve the
mean-field equations using the \emph{measured} time-dependent transport coefficients and
compare to the mean-field evolution from the simulations. This provides a thorough check 
that the measured  coefficients are correct, without relying 
on any assumptions about the type of dynamo, or simplifications to the form of $\bm{\mathcal{E}}$.

Calculations are carried out using the nonlinear
MHD equations (Eq.~\eqref{eq:MHD}),
with homogenous Cartesian geometry, periodic boundary conditions
in the azimuthal ($y$) and vertical ($z$) directions, and shearing periodic boundary conditions in the
radial ($x$) direction.  We use the {\scshape Snoopy} code \citep{Lesur:2007bh}, which applies
the Fourier pseudospectral method (in the shearing frame), and system rotation is included in some simulations through
a mean Coriolis force.
The flow field forcing ($\bm{\sigma}_{\bm{u}}$ in Eq.~\eqref{eq:MHD}) is nonhelical, 
white noise in time, isotropic, 
and centered in wavenumber space at $\left| \bm{k} \right|=6\pi$ (with width $6\pi /5$).
All simulations presented here use a box 
of size $(L_{x},L_{y},L_{z})=(1,4,2)$ with a resolution of $(N_{x},N_{y},N_{z})=(64,128,128)$,
and we take $\bar{\eta}=1/2000$ ($\mathrm{Rm}=2000$) $\mathrm{Pm}=8$ ($\mathrm{Re}=250$). To 
test convergence, we have run several cases (both with and without rotation) at twice the resolution,
and there is no discernible difference with lower resolution runs in either the spectrum, turbulence level, 
or mean-field evolution\footnote{Since there are relatively large differences between realizations 
(see Fig.~\ref{c5:fig:By examples}), we compare a variety of the lower resolution cases with the higher
resolution runs, and note that disparities between different low resolution realizations are as severe 
as those between the low and high resolution runs. We thus conclude that the differences between
the lower and higher resolutions are negligible at these parameters, and use the lower resolution for computational 
reasons in the ensemble of realizations. 
}.

Our choice of this numerical setup for the simulation ensembles is motivated both by
the calculations of \citet{Yousef:2008ix} with unstable small-scale dynamo (see their figure 9), 
and from studies of MRI turbulence in the shearing box\footnote{The 
numerical setup, aside from the noise source, is identical that of zero-net-flux unstratified accretion disk simulations.}. 
In particular, the  relatively low Reynolds 
numbers are chosen both for computational reasons (100 simulations are run for each parameter set), 
and so 
that there is no transition to self-sustaining turbulence if $\bm{\sigma}_{\bm{u}}=0$. Thus, we choose Reynolds numbers 
that are intermediate between the small-scale dynamo being stable (on the low side)
and the system transitioning to turbulence in the absence of noise (on the high side).
While
similar mechanisms may be operating  in the case of self-sustaining  turbulence \citep{Lesur:2008fn,Lesur:2008cv},
it is certainly a complicating influence that is more easily ignored for the
purposes of this study. The relatively high $\mathrm{Pm}$ is chosen 
for the obvious reason of enhancing  $\bm{b}$ in comparison to
$\bm{u}$, while still allowing for a moderate range
of scales in $\bm{u}$. It seems worth emphasizing that we do not
consider these measurements to be firm proof of the magnetic shear-current effect's 
importance at high $\mathrm{Rm}$; rather, they serve as a demonstration
that it is possible for the small-scale dynamo to significantly change $\eta_{yx}$,  and  as motivation for 
further studies at higher Reynolds numbers and with different numerical setups.

\subsection{Measurement of the transport coefficients}\label{sec:eta measurements}

In this section we describe the methods--- the test-field method \citep{Schrinner:2005jq}, and
the projection method (based on \citealt{Brandenburg:2002cia})---for 
obtaining the transport coefficients from simulations. Those 
readers who are primarily interested in results may wish to skip directly to Sec.~\ref{sec:results Rm=2000}.
 Since the projection
method is uncommon in the dynamo literature, its accuracy is verified in appendix~\ref{app:verification} 
through direct comparison to test-field
method calculations for low-Rm nonhelical shear dynamos over a range of $\eta_{yx}$. 
While the test-field method gives unambiguous
answers for kinematic transport coefficients (before the small-scale 
dynamo saturation), results can  become more difficult to interpret
in the presence of magnetic fluctuations
 \citep{Cattaneo:2009cx,Hubbard:2009dn,Rheinhardt:2010do}. In contrast,
the projection method  does not rely on any assumptions regarding the importance
of small-scale magnetic fields, operating purely from the mean-field
data from a given simulation. In addition to this method, we have
also applied a weighted least-squares method, fitting simulation data
for a single mode \citep{Kowal:2005uq}. This has led to almost identical
results for the low-$\mathrm{Rm}$ test cases and the main results given here. 
However, the least-squares method was generally found to be
somewhat less reliable and rather delicate, and we do not discuss the details. 
Another possibility for measuring transport coefficients, which could be explored 
for nonhelical shear dynamos in future work, is given in \citet{Tobias:2013bk}. 

\subsubsection{Test-field method}

The test-field method \citep{Schrinner:2005jq}, which is used for calculating transport
coefficients before small-scale dynamo saturation, 
 has become a standard tool in dynamo studies \citep{Brandenburg:2008bc}, so
we discuss this only briefly. The method involves solving for a set of $Q$ ``test fields'' $\bm{b}^{q}$ 
(where $q=1\rightarrow Q$),
in addition to the standard MHD equations. The test fields satisfy the small-scale 
induction equation,
\begin{equation}
\partial_{t}\bm{b}^{q} = \nabla\times (\bm{u}\times \bm{B}^{q}) + \nabla \times (\bm{U}\times \bm{b}^{q}) + \nabla \times \left(\bm{u}\times \bm{b}^{q} -\overline{\bm{u}\times \bm{b}^{q}}\right)+\bar{\eta} \nabla^{2}\bm{b}^{q},\label{eq:TF equations}
\end{equation}
where $\bm{B}^{q}$ are a set of $Q$ test mean fields (specified at the start of the simulation), 
and $\bm{u}$ and $\bm{U}$ are taken 
from the simulation. By calculating the EMF $\bm{\mathcal{E}}^{q}=\overline{\bm{u}\times \bm{b}^{q}}$ that 
results from a variety of $\bm{B}^{q}$, one can determine the transport coefficients. The test-field method's simplest---and most obviously meaningful---use, is to utilize a $\bm{u}$ field that is
unaffected by $\bm{b}$ or $\bm{B}$, thus calculating kinematic transport coefficients\footnote{One complication
is the small-scale dynamo of the test-fields, which can be unstable and result in
exponential growth of $\bm{b}^{q}$. This can be circumvented by reseting $\bm{b}^{q}$ 
periodically (every $T_{\mathrm{reset}}$), such that it does not become large in 
comparison to $\bm{u}$, but it is important to ensure results are independent of the choice of $T_{\mathrm{reset}}$.}.
A simple extension is the ``quasi-kinematic'' method \citep{Brandenburg:2008hc,Hubbard:2009dn}, 
for which one  runs an MHD simulation
in which $\bm{u}$ is influenced by self-consistent magnetic fields, and extracts $\bm{u}$ to insert
into the test-field equations. This 
can  most obviously be used to understand how the modification of $\bm{u}$ by $\bm{b}$ or $\bm{B}$ affects the kinematic 
coefficients (see, for example, \citealt{Gressel:2013be}), but the direct effect  of $\bm{b}$ fluctuations is not  included. 
A variety of subtleties exist, however, and care must be used in interpreting results; see \citet{Hubbard:2009dn}.

\subsubsection{Projection method}

Inclusion of the direct effect of $\bm{b}$ on transport coefficients in the test-field method 
introduces significant complications and ambiguities, primarily because it can be difficult
to ensure that the test fields $\bm{b}^{q}$ and $\bm{u}^{q}$ are linear in the test mean fields.  
A method has been proposed and explored in \citet{Rheinhardt:2010do}; however, given 
its complications and early stage of development, we choose to use the projection method 
detailed below
to calculate mean-field transport coefficients after small-scale dynamo saturation.
This method makes no assumptions regarding the importance of small-scale magnetic fluctuations, 
simply utilizing mean-field and EMF data extracted from standard MHD simulation.

The starting point of the method is the standard Taylor expansion 
of $\bm{\mathcal{E}}$  in terms of $\bm{B}$. In coordinates this is (cf.~Eq.~\eqref{c4:eq:SC sa eqs}),
\begin{subequations}
\begin{gather}
\mathcal{E}_{x}=\alpha_{xx}B_{x}+\alpha_{xy}B_{y}-\eta_{xy}\partial_{z}B_{x}+\eta_{xx}\partial_{z}B_{y}, \\
\mathcal{E}_{y}=\alpha_{yx}B_{x}+\alpha_{yy}B_{y}-\eta_{yy}\partial_{z}B_{x}+\eta_{yx}\partial_{z}B_{y}.
\end{gather}\label{c5:eq:EMF general}\end{subequations}
Note that we have not necessarily 
assumed linearity in $\bm{B}$, since $\alpha_{ij}$ and $\eta_{ij}$ are not assumed constant. 
The basic idea of the projection method, proposed in \citet{Brandenburg:2002cia},
is to extract time-series data for $\mathcal{E}_{i}$ and $B_{i}$
from nonlinear simulation, solving for the transport coefficients
in Eq.~\eqref{c5:eq:EMF general} at each time point. In principle,
all coefficients can be solved for directly, given $\bm{B}$ and
$\bm{\mathcal{E}}$ data that consists of at least 2 Fourier modes. One
calculates 
\begin{equation}
\bm{E}^{(i)}=\left(\left\langle B_{x}\mathcal{E}_{i}\right\rangle ,\left\langle B_{y}\mathcal{E}_{i}\right\rangle ,\left\langle \partial_{z}B_{x}\mathcal{E}_{i}\right\rangle ,\left\langle \partial_{z}B_{x}\mathcal{E}_{i}\right\rangle \right)^{T}
\end{equation}
and the matrix 
\begin{equation}
M=\left(\begin{array}{cccc}
\left\langle B_{x}B_{x}\right\rangle  & \left\langle B_{x}B_{y}\right\rangle  & \left\langle B_{x}\partial_{z}B_{x}\right\rangle  & \left\langle B_{x}\partial_{z}B_{x}\right\rangle \\
\left\langle B_{y}B_{x}\right\rangle  & \left\langle B_{y}B_{y}\right\rangle  & \left\langle B_{y}\partial_{z}B_{x}\right\rangle  & \left\langle B_{y}\partial_{z}B_{y}\right\rangle \\
\left\langle \partial_{z}B_{x}B_{x}\right\rangle  & \left\langle \partial_{z}B_{x}B_{y}\right\rangle  & \left\langle \partial_{z}B_{x}\partial_{z}B_{x}\right\rangle  & \left\langle \partial_{z}B_{x}\partial_{z}B_{y}\right\rangle \\
\left\langle \partial_{z}B_{y}B_{x}\right\rangle  & \left\langle \partial_{z}B_{y}B_{y}\right\rangle  & \left\langle \partial_{z}B_{y}\partial_{z}B_{x}\right\rangle  & \left\langle \partial_{z}B_{y}\partial_{z}B_{y}\right\rangle 
\end{array}\right),
\end{equation}
where $\left\langle \cdot\right\rangle $ here denotes an average
over $z$ and possibly time (the system  statistically homogenous
in $z$). Then, solving 
\begin{equation}
\bm{E}^{(i)}=M\bm{C}^{(i)},
\end{equation}
for $C^{(1)}=\left(\alpha_{xx},\alpha_{xy},-\eta_{xy},\eta_{xx}\right)$, 
$C^{(2)}=\left(\alpha_{yx},\alpha_{yy},-\eta_{yy},\eta_{yx}\right)$,
one obtains the full set of transport coefficients. 

The data for $\bm{\mathcal{E}}$ and $\bm{B}$ are generally quite noisy and 
some care is required to avoid spurious effects that lead to incorrect
results. In particular, while pure white noise in each variable will
average to zero over time, there are correlations between
components that can significantly pollute the data. 
These correlations arise from the fact that 
Eq.~\eqref{c5:eq:EMF general} is not the only expected relationship between components
of $\bm{B}$ and $\bm{\mathcal{E}}$; $\bm{B}$ is also directly 
driven by $\bm{\mathcal{E}}$, and itself, through
\begin{equation}
\partial_{t}\bm{B} = -SB_{x}\hat{\bm{y}} + \nabla\times\bm{\mathcal{E}} + \bar{\eta}\triangle \bm{B}.\label{eq:B driving}
\end{equation}
From Eq.~\eqref{eq:B driving} and by examining data, it is found that the most harmful
of the correlations are a correlation between $B_{x}$ and $B_{y}$ [as
expected due to $-SB_{x}$ in Eq.~\eqref{eq:B driving}] and a correlation
between fluctuations in $\mathcal{E}_{y}$ and $B_{x}$ 
($B_{x}$ is directly driven by $\partial_{z}\mathcal{E}_{y}$)\footnote{
The correlation between  $B_{y}$ and $\mathcal{E}_{x}$ is not so  damaging as that  between  $B_{x}$ and $\mathcal{E}_{y}$
due to the $-SB_{x}$ term in the $B_{y}$ equation and larger range
of $B_{y}$ values explored throughout a simulation.}.
Note that this correlation of $\mathcal{E}_{y}$ and $B_{x}$  is not the same as a 
nonzero $\alpha_{yx}$ or $\eta_{yy}$
coefficient. Specifically, a noisy change in the imaginary part of $\mathcal{E}_{y}$
by $\epsilon$ will cause a change in $B_{x}$ of $\sim k\epsilon\Delta t$
after some time $\Delta t$ (related to the correlation time of the $\mathcal{E}_{y}$
noise). If the noise fluctuations are of similar or larger magnitude
than the range of $B_{x}$ and $\mathcal{E}_{y}$ explored over
the course of the calculation, this correlation can cause a
\emph{negative} value for the fit parameter $\eta_{yy}$, since the {scatter} of the data 
has a preferred slope.  
In fact, a consistently negative calculated
value for $\eta_{yy}$ is the most prominent spurious effect in simulations,
which was also noted in \citet{Brandenburg:2002cia} without explanation. That this is purely a 
consequence of the projection method, and not physical, can
be established by comparison to test-field calculations (see appendix~\ref{app:verification}). 
Importantly,  the value of $\eta_{yy}$
is coupled to that of $\alpha_{iy}$ and $\eta_{yx}$. This implies one cannot simply
ignore this effect and settle with not knowing $\eta_{yy}$, since the average values of other coefficients
will also become polluted.

The basic  approach to overcoming these issues described above is  to minimize the
influence of $B_{x}$ on the calculation, to the extent possible.
This is motivated by the fact that $B_{x}$ is very noisy in comparison
to $B_{y}$, and is involved in both of the aforementioned damaging correlations.
The approach works very well for shear dynamos because $B_{x}$ is
much smaller than $B_{y}$ (e.g., in the simulations presented in this work, 
$B_{x}$ is usually between 25 and 150 times smaller than
$B_{y}$ depending on the realization). In addition, those transport
coefficients that require $B_{x}$ for their calculation (e.g., $\eta_{xy}$) are substantially
less interesting, since they do not significantly effect the dynamo
growth rate. To enable this reduction in the
influence of $B_{x}$,  two approximations are made to Eq.~\eqref{c5:eq:EMF general}.
The first and most important is to assume that diagonal transport coefficients are equal,
$\eta_{yy}=\eta_{xx}$ and $\alpha_{yy}=\alpha_{xx}$. This is not
strictly required by the symmetries of the turbulence with shear (\citealt{Radler:2006hy}; \citetalias{Analytic}),
but a variety of test-field calculations, including those after saturation
of the small-scale dynamo (i.e., quasi-kinematic calculations; see \citealt{Hubbard:2009dn,Gressel:2010dj,Gressel:2015ev}),
have shown this to be the case to a high degree of accuracy. The second approximation
is to neglect $\eta_{xy}$ and $\alpha_{yx}$. This is justified by
the fact that $B_{x}\ll B_{y}$ and $\eta_{xy}<\eta_{xx}$ on average,
thus its effect on the mean value of $\eta_{xx}$ should be very small. This approximation is 
not strictly necessary and similar results can be obtained with
$\eta_{xy}$ and $\alpha_{yx}$ included;
however, these coefficients fluctuate wildly in time (far more than $\eta_{xx}$
for example) and cause increased fluctuations in the values of the
other transport coefficients.

It is useful to briefly consider the proportional error in $\eta_{xx}$
and $\eta_{yx}$ that might arise from these approximations. First,
in considering the neglect of $\eta_{xy}$, one starts with the conservative
estimate $25B_{x}\approx B_{y}$. Noting that test-field calculations
give $\eta_{xy}\sim0.25\eta_{xx}$ for the simulations given in the
manuscript (see also \citealt{Brandenburg:2008bc}), we see that
this approximation should cause less than a $1\%$ systematic error
in $\eta_{xx}$. Second, since we are primarily interested in determining
$\eta_{yx}$, let us consider the error in $\eta_{yx}$ that results from an error
in $\eta_{yy}$ (caused by either the neglect of $\eta_{xy}$ or the
assumption $\eta_{xx}=\eta_{yy}$). Noting that $B_{x}\sim-k\sqrt{\eta_{yx}/S}B_{y}$
for a coherent shear dynamo, we can estimate that $ik\eta_{yx}B_{y}\gtrsim ik\eta_{yy}B_{x}$
when $k\eta_{yy}\lesssim\sqrt{\left|S\eta_{yx}\right|}$. This inequality
is satisfied if the coherent dynamo has a positive growth rate; thus,
very approximately, at marginality one would expect the proportional errors
in $\eta_{yx}$ and $\eta_{yy}$ to be similar. Combining
these two conclusions, one should expect the two approximations to cause very
little systematic error in the determination of $\eta_{yx}$,
despite the coefficient's small values.

To summarize the previous paragraphs, we shall fit 
\begin{subequations}
\begin{gather}
\mathcal{E}_{x}=\alpha_{yy}B_{x}+\alpha_{xy}B_{y}+\eta_{xx}\partial_{z}B_{y}, \\
\mathcal{E}_{y}=\alpha_{yy}B_{y}-\eta_{xx}\partial_{z}B_{x}+\eta_{yx}\partial_{z}B_{y},
\end{gather}\label{c5:eq:EMF spec}\end{subequations}
to simulation data at each time point. Since there are now fewer coefficients
than rows of $\bm{E}^{(i)}$, the matrix equations are solved in the
least-squares sense. One final difference from the method as utilized
in \citet{Brandenburg:2002cia} is a filtering of the data to
include only the first two Fourier modes. This is done to improve
scale separation, since the small scales of the mean field will be
dominated by fluctuations due to the finite size of the horizontal
average, and cannot be expected to conform to the ansatz in Eq.~\eqref{c5:eq:EMF general}\footnote{
This filtering violates the Reynolds averaging rules (specifically $\langle \langle g \rangle_{f} h \rangle_{f} \neq \langle g\rangle_{f}  \langle h \rangle_{f}$, where  $\langle \cdot \rangle_{f}$ is the Fourier average). While not technically required for the validity of Eq.~\eqref{c5:eq:EMF general} 
(which relies only on some level of scale separation), the rules are required for the
use of the mean-field induction equation (Eq.~\eqref{eq:B driving}) in the first place, since
$\langle \bm{u}\times \bm{B} \rangle_{f}$ may drive the Fourier-averaged field in addition to $\bm{\mathcal{E}}$ and $\bm{U}\times \bm{B}$ (this is an aliasing effect). To ensure this does not adversely affect results, we have
verified that the coefficients are essentially independent of the number of Fourier modes $n$ retained in the projection, 
up to $n\approx 5$ or $6$.}.

Finally, we note that $\alpha$ coefficients can be excluded from these
calculations altogether, and since their average over long times vanishes,
this does not affect the results for $\eta_{ij}$. We have chosen
to permit nonzero $\alpha$ in all calculations presented below, both as a consistency check 
and because over shorter
time-windows $\alpha$ may not average to exactly zero. Nonetheless,
repeating all calculations presented below and in appendix~\ref{app:verification} with
$\alpha_{ij}=0$ imposed artificially, one obtains the same results (to within the margin of error). This illustrates that in the neglect of transport coefficients considered above 
(e.g., $\alpha_{yx}$), it is only necessary to consider the errors arising from neglect of $\eta$ 
coefficients, since those due to neglect of $\alpha$ coefficients average
to zero.

\subsubsection{Verification}
To ensure the accuracy of results---especially with regards to possible systematic errors---it 
is crucial to verify the projection method. We do this with two independent approaches.
First, in appendix~\ref{app:verification}, the projection method is used to calculate kinematic transport coefficients for  low-$\mathrm{Rm}$ 
nonhelical shear dynamos, allowing a direct comparison to the 
 kinematic test-field method. The study is carried out for dynamos with a range of positive and
negative $\eta_{yx}$ by changing the rotation (see \citetalias{LowRm}), and is in a regime where the
stochastic-$\alpha$ effect is significant. This ensures that the projection method
does not inadvertently capture a property of the dynamo
growth rate, rather than the coherent transport coefficients.
Second, we verify the calculated transport coefficients are correct \emph{a posteriori} for the main simulation results  
 (Sec.~\ref{sec:numerical results}). This is done by solving the mean-field equations 
(Eq.~\eqref{c4:eq:SC sa eqs}) using the time-dependent transport coefficients $\alpha_{ij}(t)$ and $\eta_{ij}(t)$
calculated with the projection method. Comparison with the mean-field evolution 
taken directly from the simulation provides a thorough check that the transport coefficients 
are being calculated correctly, without relying on assumptions about the nature 
of the dynamo (aside from the mean-field ansatz), or the importance of approximations made to the form of the EMF (i.e., Eq.~\eqref{c5:eq:EMF spec}).

\begin{figure}
\centering{}\includegraphics[width=0.535\linewidth]{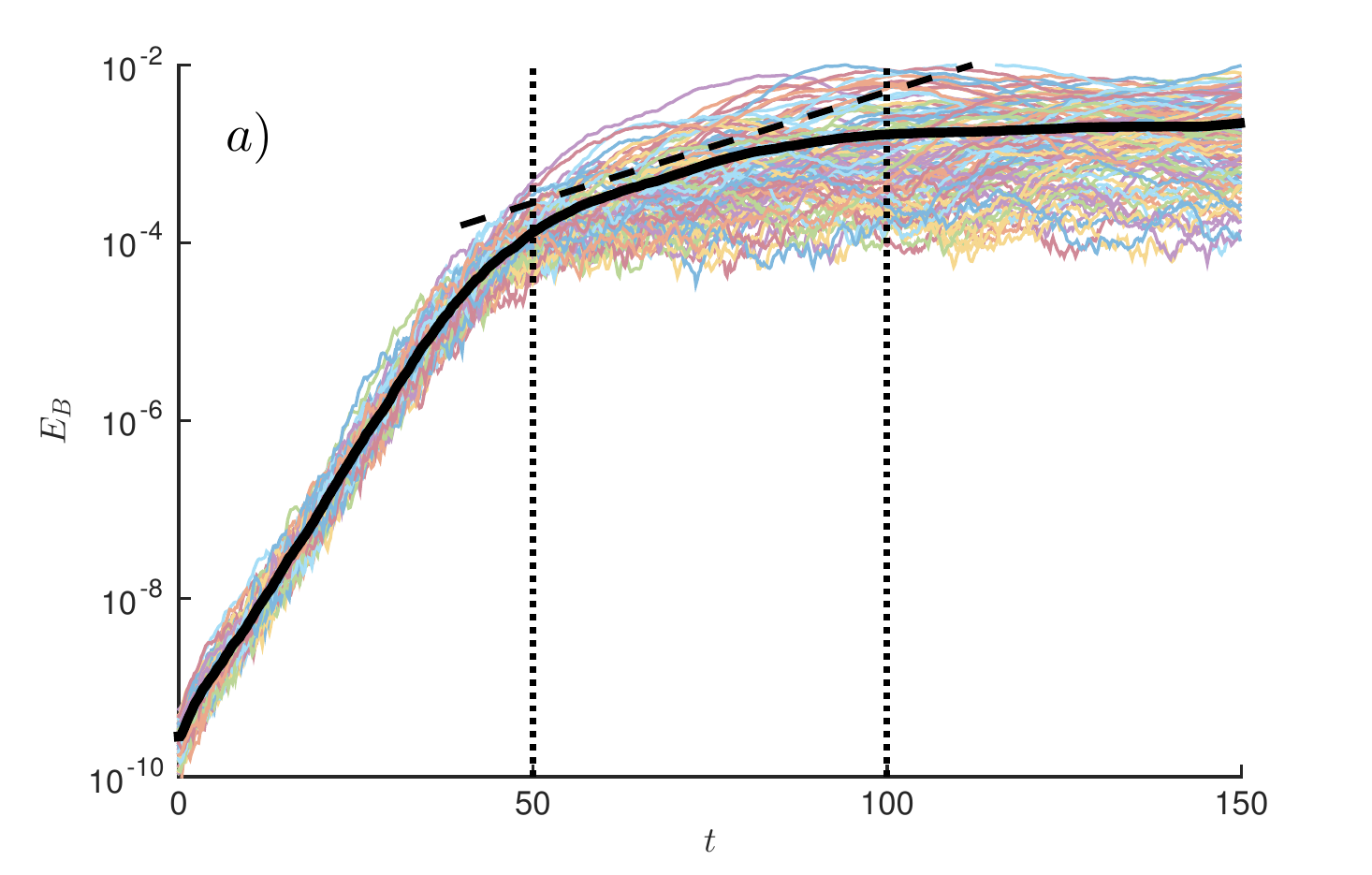}\includegraphics[width=0.465\linewidth]{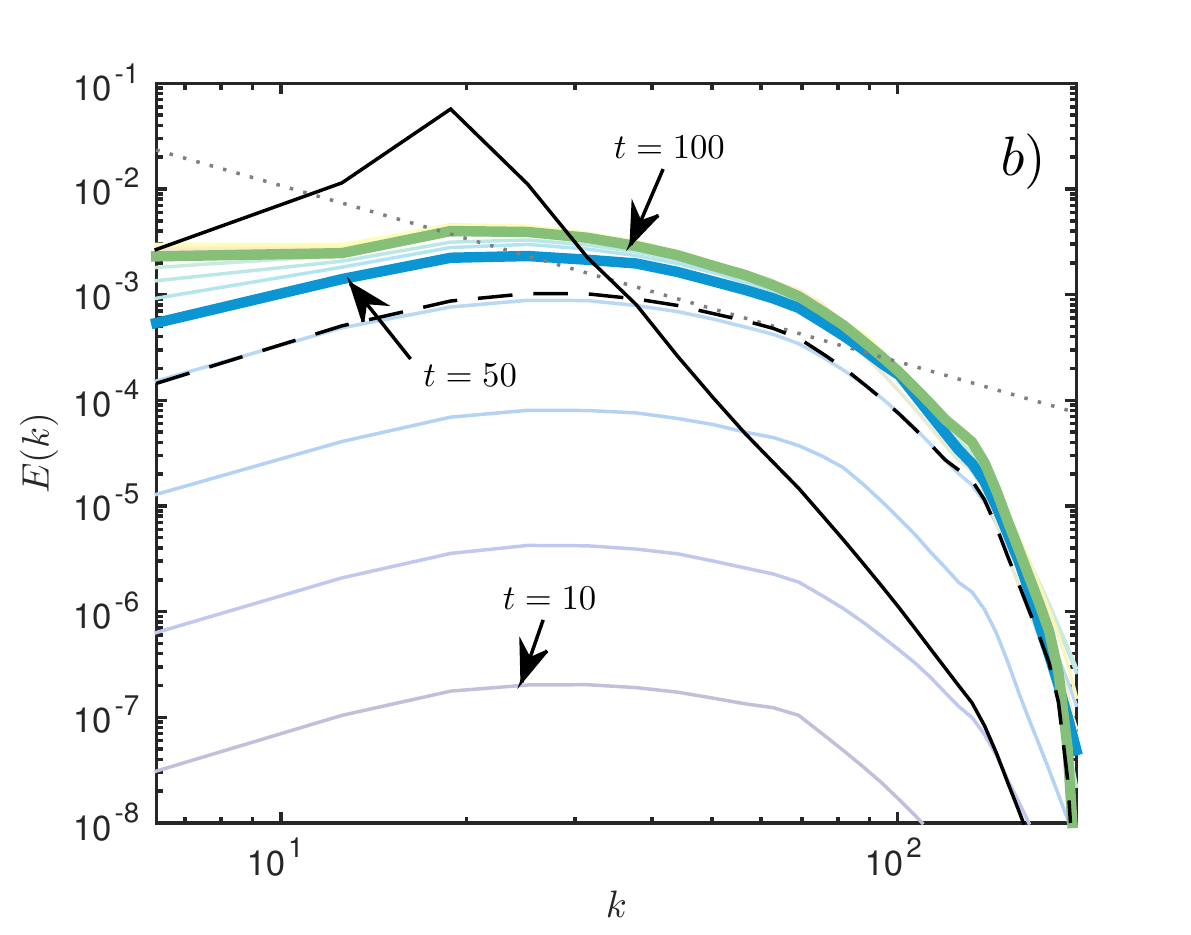}\caption{(a) Time development of the average mean-field energy $E_{B}=L_{z}^{-1}\int dz\, \bm{B}^{2}$ for the rotating simulation set. Each faded color curve illustrates a single realization, while the 
thick black curve shows the mean over all realizations. The dotted vertical lines indicate the saturation of
the small-scale dynamo and the nonlinear saturation of the large-scale dynamo (the projection method is applied between these), while the dashed line is simply an approximate fit to the large-scale dynamo growth phase. (b) Shell-averaged turbulent spectra (of $\bm{B}_{T}$ and $\bm{U}_{T}$) for the rotating simulations shown in (a). The colored lines (from blue to yellow) illustrate the growth of the magnetic spectrum in time (averaged over all simulations), with the spectra at $t=50$, $t=100$, and $t=150$ highlighted by thicker 
lines. The solid black line illustrates the velocity spectrum (peaked at $k\approx 6\pi$), while the dashed line 
shows the magnetic spectrum from an identical simulation, but without velocity shear ($S=0$), and averaged in time from  $t=50$ to $t=150$. Evidently, as also seen in \citet{Yousef:2008ix}, the second period
of large-scale field growth is absent when $S=0$ (note that the velocity spectrum is essentially identical to the case with velocity shear). The dotted line simply illustrates a $k^{-5/3}$ spectrum for the sake of clarity. The slight bump in the spectrum at high $k$ is caused by spectral reflection from the 
grid cutoff; however, since this is well into the exponential fall-off and at very low energy we are confident that this does not affect  large-scale evolution (note that the spectrum in the few double resolution
 simulations is essentially identical, aside at and above the bump itself). 
\label{fig:EandSpectrum}}
\end{figure}

\subsection{Numerical results---the magnetically driven dynamo}\label{sec:results Rm=2000}\label{sec:numerical results}

In this section, we show that small-scale fields arising self-consistently through
the small-scale dynamo can drive a coherent large-scale dynamo.
To this end, we apply the methods discussed in the previous section to 
calculate transport coefficients before and after 
the saturation of the small-scale dynamo. 
The technique is applied to ensembles of 100 simulations, both with and without Keplerian
rotation.

\begin{figure}
\centering{}\includegraphics[width=0.8\linewidth]{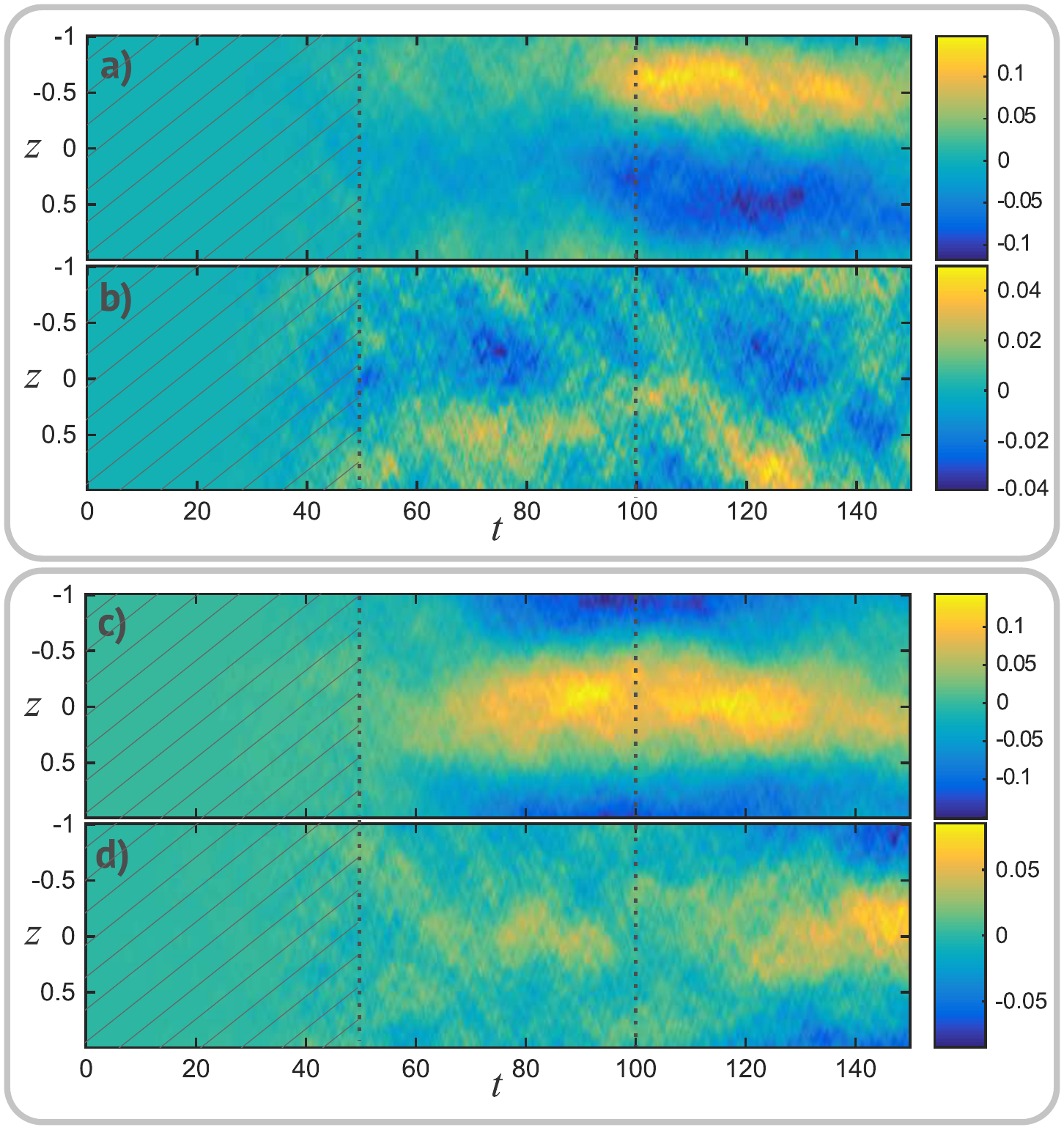}\caption{
Example spatiotemporal $B_{y}$ evolutions for (a-b) non-rotating ($\Omega=0$), 
and  (c-d) Keplerian rotating ($\Omega=2/3$) driven turbulence (parameters described in the text).
The first examples in each case {(}(a) and (c){)} show $B_{y}$ when
a coherent dynamo develops, while the second examples {(}(b) and (d){)}
illustrate the case when it is more incoherent. The main factors in
distinguishing these are the coherency in phase of $B_{y}$ over some
time period and the amplitude at saturation,
which is larger in the coherent cases. In general the rotating simulations
are substantially more coherent. The hatched area illustrates the
region of small-scale dynamo growth. The projection method used to compute
transport coefficients (see Fig.~\ref{c5:fig:Transport coeffs}) is
applied between the dashed lines ($t=50\rightarrow100$).\label{c5:fig:By examples} }
\end{figure}

Before continuing, we demonstrate that there is indeed a large-scale dynamo that
develops after saturation of the small-scale dynamo. This is shown both in  Fig.~\ref{fig:EandSpectrum}, which gives the time development of the mean-field energy and turbulent spectra, and in Fig.~\ref{c5:fig:By examples}, which illustrates the spatiotemporal evolution 
of $B_{y}(z,t)$ in several example realizations. From Fig.~\ref{fig:EandSpectrum}, we clearly see 
the fast growth of the small-scale dynamo until its saturation at $t\approx 50$. (This 
is observable in  Fig.~\ref{fig:EandSpectrum}(a), which shows only the mean-field, because $\bm{B}$
is in approximate equipartition with the small-scales due to the finite domain.) Following this there is a slower period of 
growth in only the largest scales of the box (this is $k_{1z}=\pi$, a factor of six less than the forcing scale), with this saturating around $t=100$ on average.
Importantly, this second period of slower growth in the mean-field is not present
without the mean shear (see Fig.~\ref{fig:EandSpectrum}(b), dashed line), despite the velocity spectrum 
being essentially identical. This illustrates that the shear causes large-scale field generation after
saturation, as also noted in \citet{Yousef:2008ix},

As shown in Fig.~\ref{fig:EandSpectrum}(a) and  Fig.~\ref{c5:fig:By examples}, 
at these parameters, the prevalence of a coherent large-scale
dynamo after saturation of the small-scale dynamo varies significantly between realizations.
Specifically, it appears that the coherent effect cannot always overcome fluctuations
in $\bm{\mathcal{E}}$ immediately after small-scale saturation, although
the dynamo always develops after a sufficiently long time {[}e.g.,
Fig.~\ref{c5:fig:By examples}(d) near $t=150${]}. This behavior
seems generic when the coherent dynamo is close to its threshold for
excitation, and similar structures were  observed at lower Rm in \citetalias{LowRm}, where
forcing in the induction equation was used to create an homogeneous bath of magnetic fluctuations. 
Nonetheless, despite this variability
in the dynamo's qualitative behavior, measurement of the transport
coefficients over the ensemble of simulations illustrates a significant decrease in $\eta_{yx}$ 
after the magnetic fluctuations reach approximate equipartition with
velocity fluctuations at small scales. 

At low times, before $t\approx 50$, the kinematic
$\alpha$ and $\eta$ are measured using the 
test-field method, fixing the mean field and calculating $\bm{\mathcal{E}}$,
with no Lorentz force \citep{Brandenburg:2005kla,Brandenburg:2008bc}.
Calculations are run from $t=0\rightarrow 2000$ with the errors estimated
through the standard deviation of the mean over 100 segments. 
Since the small-scale dynamo grows quickly, test fields are reset
every $t=5$. After small-scale saturation, we utilize the projection method 
(Sec.~\ref{sec:eta measurements}) to measure coefficients
directly from the observed mean-field and EMF evolution\footnote{
While it would be ideal to measure coefficients before 
saturation using the projection method  for consistency, this is difficult. 
In particular, the method  is hindered by the small-scale dynamo causing
the mean-field evolution to be completely overwhelmed by small-scale noise. 
We explored the possibility of seeding initial conditions with large-scale fields
to obtain a short period of kinematic evolution; however, results were inconclusive due to very 
high levels of noise in the measurements. }. 
The time window of these measurements has been limited to 
$t=50\rightarrow100$, since growth is seen to stop at $t\approx100$ in many realizations (see Fig.~\ref{fig:EandSpectrum} and Figs.~\ref{c5:fig:By examples}(a) and (c)).
Since this saturation presumably occurs due to a nonlinear change in the transport coefficients at large
$\bm{B}$ (e.g., a change in sign of $\eta_{yx}$), it is important to not include this saturation
phase in the measurement of $\eta_{yx}$.
As should be expected from Fig.~\ref{c5:fig:By examples} and due to the short time window, 
measurements of the transport coefficients after small-scale saturation vary significantly between realization.
Nonetheless, an average over the ensemble illustrates a statistically
significant change in $\eta_{yx}$ that is consistent with observed
behavior, in both the rotating and non-rotating simulation ensembles.

\begin{figure}
\begin{centering}
\includegraphics[width=0.9\linewidth]{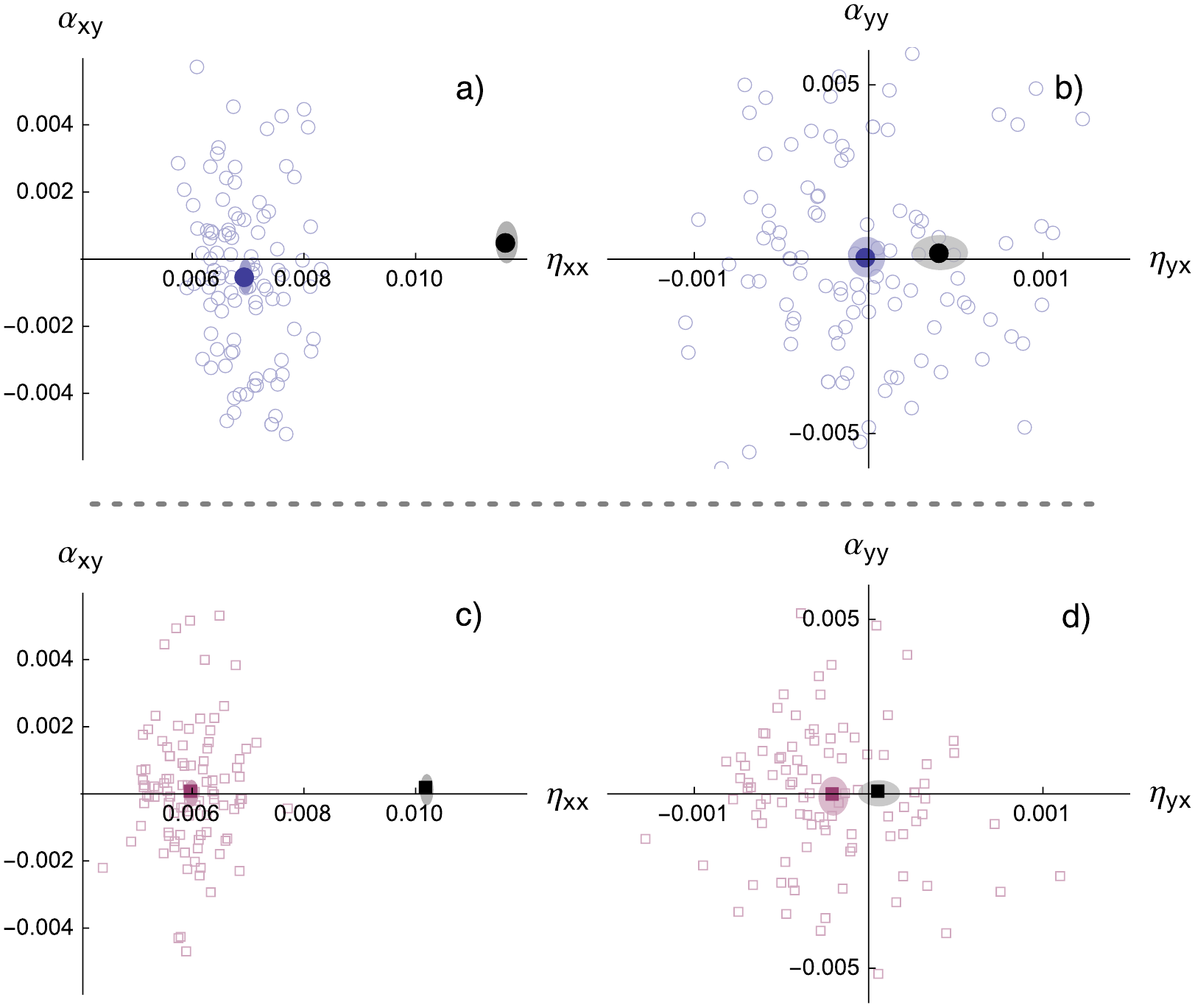}
\par\end{centering}
\centering{}\caption{
Measurements of the turbulent transport coefficients for 100 realizations
of the simulations in Fig.~\ref{c5:fig:By examples};
(a) $\eta_{xx}$ coefficients, no rotation, (b) $\eta_{yx}$ coefficients,
no rotation, (c) $\eta_{xx}$ coefficients, rotating, (d) $\eta_{yx}$
coefficients, rotating. Unfilled markers in each plot (circles and
squares for non-rotating and rotating runs respectively) show coefficients
measured from each of the individual realizations, with mean values
displayed by solid markers and the shaded regions indicating error
in the mean (2 standard deviations). Black markers illustrate
the kinematic transport coefficients, with grey shaded regions indicating
the error. After saturation of the small-scale dynamo, 
$\eta_{ij}$ is calculated
using the projection method, taking the mean from
$t=50$ to $t=100$. This limited time window is chosen to avoid capturing
the saturation phase of the large-scale dynamo, since $\eta_{ij}$
is presumably modified in this phase. In both methods used to compute
transport coefficients,  the corresponding
$\alpha$ coefficients are also calculated. In all cases these are zero to within error
as expected, and the scatter between simulations is of a similar magnitude
to that of $\eta_{ij}$ if their different units are accounted for
(it is necessary to divide $\alpha$ by a characteristic $k$ value).
\label{c5:fig:Transport coeffs}}
\end{figure}

Figure~\ref{c5:fig:Transport coeffs} illustrates the results. In
the kinematic phase without rotation, we see $\eta_{yx}=\left(4.1\pm1.6\right)\times10^{-4}$,
in qualitative agreement with previous studies \citep{Brandenburg:2008bc}.
With rotation, we find $\eta_{yx}=\left(0.6\pm1.2\right)\times10^{-4}$,
consistent with a reduction in $\eta_{yx}$ due to the $\bm{\Omega}\times\bm{J}$
effect (\citealt{Krause:1980vr}, but note the deviation from the lower-$\mathrm{Rm}$
case and SOCA result, which predicts negative $\eta_{yx}$). After saturation of the small-scale dynamo,
$\eta_{yx}=(-0.1\pm1.0)\times10^{-4}$ for the non-rotating case,
while $\eta_{yx}\approx-\left(2.0\pm0.8\right)\times10^{-4}$ in the
rotating case. The reduction of each is the same to within error. Values
for the diagonal resistivity are smaller after saturation, which is consistent
with the observed decrease in the velocity fluctuation energy (by a factor $\sim1.4$).

The numerical values of $\eta_{xx} $ and $\eta_{yx}$ show that the coherent dynamo
is slightly stable on average in the non-rotating case and marginal
in the rotating case. However, the coefficients vary significantly
between realizations, sometimes yielding larger growth rates, 
and it is important to check that the observed mean-field evolution has 
some relation to this variation. This serves two purposes. First, it acts 
as a check that the projection method is measuring the transport coefficients correctly.
Second, it illustrates that those realizations exhibiting the strongest 
growth are indeed being driven by the shear-current mechanism; that is,
they are driven by $\eta_{yx}$ rather than residual variation of $\alpha$ about mean zero.
This corroborates the earlier conclusion that the
approximately constant phase of $B_{y}(z,t)$ in the development of the
dynamo (Fig.~\ref{c5:fig:By examples}) is inconsistent with an $\alpha$ effect.

\begin{figure}
\includegraphics[width=1\textwidth]{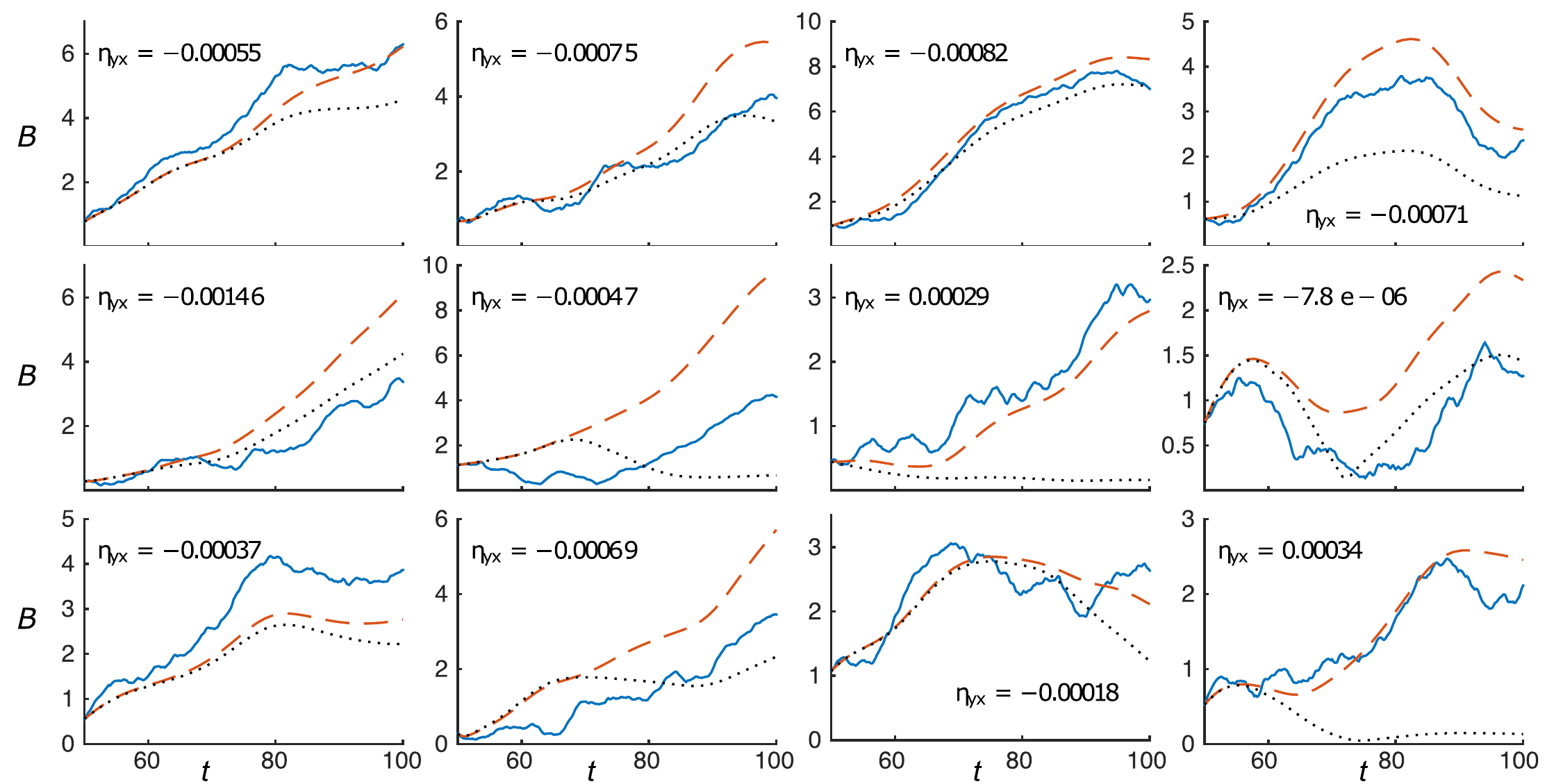}
\caption{Evolution of the mean-field magnitude for the first 12 of the ensemble
of rotating simulations discussed in the manuscript. Here $B(\vert\hat{B}_{x}^{1}\vert^{2}+\vert\hat{B}_{y}^{1}\vert^{2})^{1/2}$ is the
mean-field magnitude,
where $\hat{B}_{i}^{1}$ is the largest scale Fourier mode of $B_{i}$.
In each plot the solid blue curve shows data taken from the simulation.
The dashed red curve shows the corresponding expected evolution, using
the smoothed calculated values of the transport coefficients (see
text). Finally, the dotted black curve illustrates the expected evolution,
artificially setting all $\alpha$ coefficients to zero. We list the
measured mean of $\eta_{yx}$ in each plot to show that higher absolute values (i.e, more negative values)
do generally lead to substantially more growth of the mean field as
expected for a coherent dynamo. For reference, at the measured $\eta_{xx}\approx0.006$,
the coherent dynamo is unstable below $\eta_{yx}=-0.00036$. \label{c5:fig:Rot evolutions}}
\end{figure}

As stated previously, the method for checking this consistency is to use the \emph{measured} 
transport coefficients to
solve for the expected evolution of the largest Fourier mode of $B_{i}$
{(}using Eq.~\eqref{c4:eq:SC sa eqs}{)}, comparing this to the
observed evolution from the full simulation. Note that we use
the time-dependent coefficients $\alpha_{ij}(t)$ and $\eta_{ij}(t)$, rather 
than the time average that is shown in Fig.~\ref{c5:fig:Transport coeffs}, 
since this provides much more information about the details of the evolution. The check is carried out for
each realization separately, initializing using the mean-field data
and filtering transport coefficients in time with a Gaussian filter
of width $5$ to remove the rapid fluctuations. Results from the first
12 realizations for rotating runs (chosen since the dynamo is stronger than in the nonrotating cases) 
are shown in 
Figs.~\ref{c5:fig:Rot evolutions}.
The agreement is generally good, with qualitatively similar features
between calculated and measured evolution in all realizations, and
many cases showing quantitative agreement. It seems that in most instances
for which there is a substantial divergence between the predicted and
observed mean-field evolution, it is due to a slight error building
up in $B_{x}$ that subsequently gets amplified enormously due to
the $-SB_{x}$ term in the $B_{y}$ equation. 

In addition to solving for expected evolution using both $\eta_{ij}(t)$
and $\alpha_{ij}(t)$ measurements, we  present calculations obtained
in an identical way, but with $\alpha_{ij}(t)$ coefficients artificially
set to zero. The purpose of this analysis is to examine
the degree to which the dynamo is driven by $\eta_{yx}$, rather than
variation in $\alpha$ about its mean of zero. Through a comparison of  the curves
with and without $\alpha_{ij}$ it is clear that in many realizations
of the rotating simulation set, the dynamo is primarily driven by
$\eta_{yx}$, as shown by the agreement between dashed and dotted
curves. Furthermore, the mean of $\eta_{yx}$ over the time interval
(printed on each subfigure; these are taken from Fig.~\ref{c5:fig:Transport coeffs}) agrees nicely with the
observed behavior. That is, large negative values for $\eta_{yx}$ correspond
to those realizations with both strong dynamo growth and good agreement
between evolution with and without $\alpha$. In contrast, realizations with lower absolute values of $\eta_{yx}$ (i.e., values for which the dynamo is stable) either grow
very little or diverge substantially between evolution with and without
$\alpha$. This shows that sometimes, for realizations in which the
magnetic shear-current effect is weaker,  a stochastic $\alpha$ effect is the primary
driver. A similar examination of the non-rotating case shows that coherent dynamo growth
is much less prevalent. In particular, while the agreement between the true  and 
calculated evolution is satisfactory (similar to Fig.~\ref{c5:fig:Rot evolutions}), 
there is generally much less mean-field growth and larger differences
with calculations for which $\alpha_{ij}$ is artificially set to zero. Since in most realizations $\eta_{yx}$ is
larger than the threshold at which the coherent dynamo becomes unstable even after 
the decrease due to magnetic fluctuations, this
is not surprising.

We thus conclude that small-scale
magnetic fluctuations act to make $\eta_{yx}$ \emph{more negative}, and that
in some realizations (or after a sufficiently long time period) a
coherent large-scale dynamo develops as a result. 
This demonstrates that  magnetic fluctuations, excited
by small-scale dynamo action, can drive large-scale magnetic field
generation. The consistency of the numerical simulations with theoretical expectations, 
as well as the general agreement of measured transport coefficients with observed 
mean-field evolution, give us confidence that the observed large-scale dynamo
is indeed a coherent effect. The mechanism is the magnetic shear-current effect, 
 arising through the contribution of magnetic fluctuations to the off-diagonal turbulent resistivity $\eta_{yx}$
in the presence of large-scale shear flow.

The most significant limitation of the studies presented in this section
is the relatively low Reynolds numbers, which were chosen to be slightly below the
transition to self-sustaining turbulence (in the absence of driving noise), as well 
as for computational reasons (since an ensemble of simulations was required). 
Specifically, the dynamo is likely  far from any asymptotic regime at high $\mathrm{Re}$ and $\mathrm{Rm}$. 
This is almost certainly true for both the small-scale dynamo and its saturation\footnote{The
critical $\mathrm{Rm}$ for onset of the small-scale dynamo with this forcing is $\mathrm{Rm}_{c}\approx 1100$, 
so we are well away from where one might expect the growth rate or saturated field level to converge \citep{Tobias:2015jp}.}, as well 
as for the large-scale magnetic shear-current effect itself.
Further, the diffusion time scale of the large-scale modes is $t\approx(\bar{\eta}(2\pi/L_{z})^{2})\approx 200$, which 
is only an order of magnitude different from the time scale of growth of the large-scale field ($t\approx30$, as can be seen from Fig.~\ref{c5:fig:Rot evolutions}). The same is true of
the separation between the turbulent forcing scale $k_{f}\approx 19$, and that of the large-scale field $k_{1}=2 \pi/L_{z} = \pi$. 
While such limitations are hardly unique in the dynamo literature, it is 
obviously pertinent to undertake future studies at much higher resolutions. Further discussion 
of some of the difficulties involved with truly understanding the astrophysical relevance of 
the magnetic shear-current effect is given in the next section.

\section{Discussion and conclusions}\label{sec:discussion}


This paper has revolved around exploration of the ``magnetic shear-current effect'' as a viable mechanism
to drive large-scale dynamos in nonhelical shear flows.  The suggestion is that a bath of homogeneous nonhelical \emph{magnetic} fluctuations, influenced by the velocity shear,
can cause a dynamo instability through an off-diagonal turbulent resistivity, 
even if there is no $\alpha$ effect.
More specifically, in response to a large-scale azimuthal magnetic field $B_{y}$,
a bath of magnetic fluctuations will produce an azimuthal electromotive force $\mathcal{E}_{y}$, proportional
to $\partial_{z}B_{y}$.  
This $\mathcal{E}_{y}$ causes the generation of a radial magnetic field,
which in turn amplifies the azimuthal field through stretching by the mean flow (the $\Omega$~effect), 
resulting in a dynamo instability.  
The effect  rests crucially on the sign of the proportionality between $\mathcal{E}_{y}$ and $\partial_{z}B_{y}$ (termed 
$\eta_{yx}$)---if the product $\eta_{yx}(\nabla\times \bm{U})_{z}$ is negative, the induced 
radial field will act to damp, rather than amplify, the azimuthal field. 

The physical picture for the magnetic shear-current effect---how magnetic fluctuations 
can interact with velocity shear 
and a large-scale field gradient to produce an $\bm{\mathcal{E}}$ of the required direction---is somewhat different from dynamo mechanisms described in previous literature (see, 
for example, \citealt{Brandenburg:2005kla,Yokoi:2013di}). In particular,
it relies on the \emph{pressure response} of the fluid 
to the Maxwell stress $\bm{B}_{T}\cdot \nabla \bm{B}_{T}$. The
basic effect arises because $\bm{b}\cdot \nabla \bm{B}$ 
creates $\bm{B}$-directed velocity perturbations from magnetic 
perturbations in the direction of magnetic shear (i.e., the $\nabla $ direction). This implies
that any variation of the $\bm{b}$ perturbation along the $\bm{B}$ direction
will create a velocity perturbation with nonzero divergence, leading to a significant pressure response.
Without velocity shear, the response is well known and fundamental for turbulent diffusion; it exactly 
cancels another term and causes the contribution to the turbulent mean-field resistivity 
from magnetic fluctuations
to vanish (this is also known as the absence of $\beta$-quenching; \citealt{Gruzinov:1994ea}). In the presence of velocity shear, 
a secondary pressure response, arising due to the stretching of the primary pressure response by the mean shear, causes perpendicular
velocity fluctuations that are correlated with the original magnetic fluctuations. The resulting EMF 
is in the required direction to generate a $\bm{B}$ that is stretched by the shear flow, enhancing the
mean field that caused the effect in the first place. 
 Thus, a mean-field dynamo 
instability can ensue at sufficiently long wavelength.

Why is magnetic shear-current mechanism interesting? We would like to give two answers
to this question: the first relates generally to dynamo theory, the second to the specific case of the  dynamo seen in simulations of
turbulence in accretion disks (the MRI dynamo).
\begin{description}
\item[General mean-field dynamo theory.~]{Much of mean-field 
dynamo theory in recent years has focused on the issue of $\alpha$ quenching \citep{Kulsrud:1992ej,Gruzinov:1994ea}. This is 
specifically related to the adverse influence of small-scale magnetic fields on large-scale 
dynamo action. Since small-scale dynamos grow faster than large-scale fields 
above moderate Reynolds numbers, large-scale dynamos may \emph{always} have to grow 
on a bath of small-scale magnetic fluctuations (\citealt{Cattaneo:2009cx}, but see also \citealt{Tobias:2014ek}).
With this in mind, the magnetic shear-current effect is the first suggestion (of which we are aware) 
for a large-scale dynamo driven by small-scale magnetic fluctuations (although quenching of the turbulent resistivity can 
lead to a dynamo with spatial variation of transport coefficients; \citealt{Parker:1993kd, Tobias:1996eu})\footnote{
The magnetic $\alpha$ effect can certainly drive a mean field  in isolation. The 
key point is that its sign is opposite to that of the kinetic 
effect, and the small-scale dynamo grows such that the two effects cancel. 
While it may be possible that instabilities 
would cause a magnetic $\alpha$ effect to overwhelm the kinematic one (for instance, the MRI in the presence of stratification, \citealt{Gressel:2010dj,Park:2012eg}), this remains
unclear. In contrast, the magnetic shear-current effect has a fixed sign, arising 
from the nonhelical part of the fluctuations.}.
Thus, in some sense, the effect is the \emph{inverse} of dynamo quenching; rather than magnetic
fluctuations overwhelming a desirable kinematic effect,  mean-field growth 
starts after small-scale dynamo saturation, driven by the small-scale field itself. 
In this work, we have given an example of this interesting behavior through targeted 
numerical experiments. These illustrate that the magnetic fluctuations resulting from 
saturation of the small-scale
dynamo cause a significant decrease (and in some cases, a  sign change) of the crucial  $\eta_{yx}$
transport coefficient, which can in turn drive a large-scale dynamo. 
Study of such magnetic dynamos in direct numerical simulations is confounded by the very short 
period of exponential growth that can be observed (in contrast with kinematic shear dynamos; \citealt{Yousef:2008ie}; \citetalias{LowRm}), and
more work is needed to better assess regimes where the effect might be dominant, or
even if it continues to operate at very high Reynolds numbers. Nonetheless, it is an interesting possibility
that may find application across a wide variety of astrophysical objects. 
}
\item[The~MRI~dynamo.~]{The central regions of accretion disks are both unstratified and lack a source 
of net kinetic or magnetic helicity, implying that an $\alpha$ effect is not possible. 
In addition, a variety of authors have found from 
simulation and theory
that the crucial  $\eta_{yx}$ is of the wrong sign for a kinematic nonhelical shear dynamo (\citealt{Radler:2006hy,Rudiger:2006gx,SINGH:2011ho}; \citetalias{LowRm}).
What then is the cause of the apparent large-scale dynamo seen in simulations? While there
is the possibility that it is driven by fluctuations in the $\alpha$ coefficients \citep{Vishniac:1997jo,Vishniac:2009il}, we would
argue that the magnetic shear-current effect is a more likely candidate: MRI simulations exhibit 
stronger magnetic than kinetic fluctuations, the Keplerian 
rotation is favorable for dynamo growth, the velocity shear is obviously important, and the nonlinear
behavior of the effect bears strong similarities to mean-field dynamics in unstratified MRI simulations. In 
addition, the basic importance of $\eta_{yx}$ in the MRI dynamo has been concluded from 
nonlinear simulation \citep{Lesur:2008cv} and perturbative calculations of the evolution of MRI modes \citep{Lesur:2008fn}.
Our suggestion that small-scale magnetic fields are in fact the primary driver thus
ties together formal mean-field dynamo theory with these studies and explains 
the special importance of strong magnetic fluctuations in MRI turbulence and dynamo.   }
\end{description}

Some of the most compelling evidence that the magnetic shear-current effect is indeed
responsible for the unstratified MRI dynamo comes from statistical simulation of the
saturation of MRI turbulence \citep{Squire:2015fk}. Statistical simulation (\citealt{Farrell:2012jm,Tobias:2011cn}) 
involves formulating equations  for \emph{statistics} of the small-scale fields $\bm{u}$ and $\bm{b}$
in the mean fields ($\bm{U}$ and $\bm{B}$), and solving these, rather than  a single turbulent realization. 
Importantly for the shear dynamo, this completely eliminates the possibility of a stochastic-$\alpha$ effect, since the $\bm{\mathcal{E}}$ that drives $\bm{B}$ is calculated directly from fluctuation statistics. Coupled with the fact that the kinematic effect is too weak to explain the dynamo \citepalias{LowRm}, it is clear that the magnetic shear-current effect is the \emph{only possible} field generation mechanism
in these calculations. Despite this, the agreement with nonlinear simulation is very good (see Fig.~2 of \citealt{Squire:2015fk}). 
Most important is the observed strong increase in the saturated mean $\bm{B}$ field, and consequently in the 
turbulent angular momentum transport, as the magnetic Prandtl
number  is increased at fixed $\mathrm{Re}$. This counterintuitive trend 
has been the source of much discussion in the MRI turbulence literature (see, for example, \citealt{Lesur:2007bh,Fromang:2007cy,Meheut:2015it}). The considerations above illustrate that it is, at least in part, 
a consequence of the $\mathrm{Pm}$ dependence of the saturation of the magnetic shear-current effect.

Looking past the unstratified MRI dynamo, we might wonder about other applications of the
magnetic shear-current effect. Large-scale velocity shear is inescapable in the universe due to the influence of gravity, while
the generic instability of small-scale dynamo at large Reynolds numbers implies that plasma turbulence
should always be accompanied by small-scale magnetic fluctuations
in near equipartition with velocity fluctuations \citep{Schekochihin:2007fy}.
However, the simulations discussed in Sec.~\ref{sec:numerical evidence}
are intended to illustrate that the magnetic shear-current effect is \emph{possible}, not necessarily that it should be 
important in every situation. Unfortunately, estimating the relevance of the effect in astrophysical scenarios in any detail requires   
more knowledge about its dependence on physical parameters---particularly the 
Reynolds numbers (and magnetic Prandtl number). There are numerous complicating factors that will arise in estimating 
these dependencies. Most obvious is the variation of transport coefficients themselves (especially $\eta_{yx}$) for a 
given level of magnetic fluctuations. While it is certainly encouraging that a variety of different methods
agree on the sign of $\eta_{yx}$, most results are truly valid only at  low Reynolds numbers\footnote{The spectral $\tau$~approximation (which predicts $(\eta_{yx})_{b}<0$ and $\left|(\eta_{yx})_{b}\right| \gg \left|(\eta_{yx})_{u}\right|$; \citealt{Rogachevskii:2004cx}) is nominally valid at high Reynolds number, but its accuracy and reliability remain unclear (see, for example, \citealt{Radler:2007hp}).}. More subtly, the relevance of the effect could depend significantly on the saturation 
level of the small-scale dynamo, which would be especially important if the kinematic shear-current
effect has the incorrect sign for dynamo action ($(\eta_{yx})_{u}$ may change sign with Reynolds number; see \citealt{Brandenburg:2008bc}). This saturation level presumably depends on $\mathrm{Pm}$, 
but may also change under the influence of velocity shear (at least at the larger of the small scales), an effect that may become significant only at very high Reynolds numbers \citep{Tobias:2014ek,Cattaneo:2014jg}. Finally, magnetic helicity and its transport are a cornerstone of modern dynamo theory \citep{Vishniac:2001wo,Field:2002fn}, but have  not
been explored in our work thus far due to the focus on the linear phases of the dynamo instability. 
Such effects will be important to consider in future studies of the saturation and nonlinear evolution of magnetic shear-current dynamos  \citep{Rogachevskii:2006hy}.
 Overall, given the general difficulty of even measuring growth rates for magnetically driven large-scale dynamos, it seems that
the magnetic shear-current effect will provide a variety of rich and interesting avenues for future exploration.

\acknowledgements
The authors would like to thank J.~Krommes, J.~Goodman, H.~Ji, G.~Hammett, and A. Schekochihin  for enlightening discussion and useful suggestions, as well as G.~Lesur for distribution of 
the  {\scshape Snoopy} code.
JS acknowledges the generous support of a Burke Fellowship and the Sherman Fairchild Foundation at Caltech, as well as a Procter Fellowship at Princeton University. This work was funded by U.S. Department of Energy Grant No. DE-AC02-09-CH11466 and computations were carried out on the Dawson  cluster
at PPPL. 

\appendix
\section{Verification of the projection method: low-$\mathrm{Rm}$ shear dynamo}\label{app:verification}
In this appendix, we verify that the projection method discussed in Sec.~\ref{sec:eta measurements} recovers the correct transport coefficients
for low-$\mathrm{Rm}$ shear dynamos, similar to those studied in \citetalias{LowRm}
and  \citet{Yousef:2008ix,Yousef:2008ie}.
The primary advantage of testing the method in this parameter regime
is that there is no small-scale dynamo and simulations
exhibit a very long kinematic growth period over which the small-scale
velocity field is unaffected by the magnetic field. It is thus straightforward
to compare results obtained with the projection method to those using
the test-field method, where the only the fluctuating part of the induction equation is solved. 

The simulations are carried out in the same numerical 
setup as used in the main text, but at $\mathrm{Re}=\mathrm{Rm}=100$
in shearing boxes of dimension $\left(L_{x},L_{y},L_{z}\right)=\left(1,1,8\right)$. The large
$L_{z}$ is chosen to allow for a long mean-field wavelength and thus enhance 
the dynamo  instability \citep{Yousef:2008ix}.
The velocity field is forced
at $k=6\pi$ to a level $u_{rms}\approx0.8$, using the same 
nonhelical $\bm{\sigma}_{\bm{u}}$ as detailed in the main text. Keeping $S=2$, we present
cases that are non-rotating, $\Omega=0$, as well as $\Omega=4/3$
(Keplerian), and $\Omega=4$, with the rotation added through the mean
Coriolis force. As discussed in \citetalias{LowRm}, this change in $\Omega$ 
causes $\eta_{yx}$ to change
sign due to the $\Omega\times J$ (or R{\"a}dler) effect. We have run 10 simulations at each parameter
set from $t=0$ to $t=1000,$ although the rotating cases saturate
earlier ($\Omega=4/3$ at $t\approx900$, $\Omega=4$ at $t\approx500$)
due to faster dynamo growth. Note that the ratio of $B_{y}$ to $B_{x}$ in 
these simulations ($\sim 10\rightarrow 30$ during growth)
 is somewhat higher  than that for the magnetically driven dynamos studied in Sec.~\ref{sec:numerical results}; thus, if anything, one might expect larger systematic errors in these simulations than the estimates
given in Sec.~\ref{sec:eta measurements}.

\begin{figure}
\begin{centering}
\includegraphics[width=0.47\textwidth]{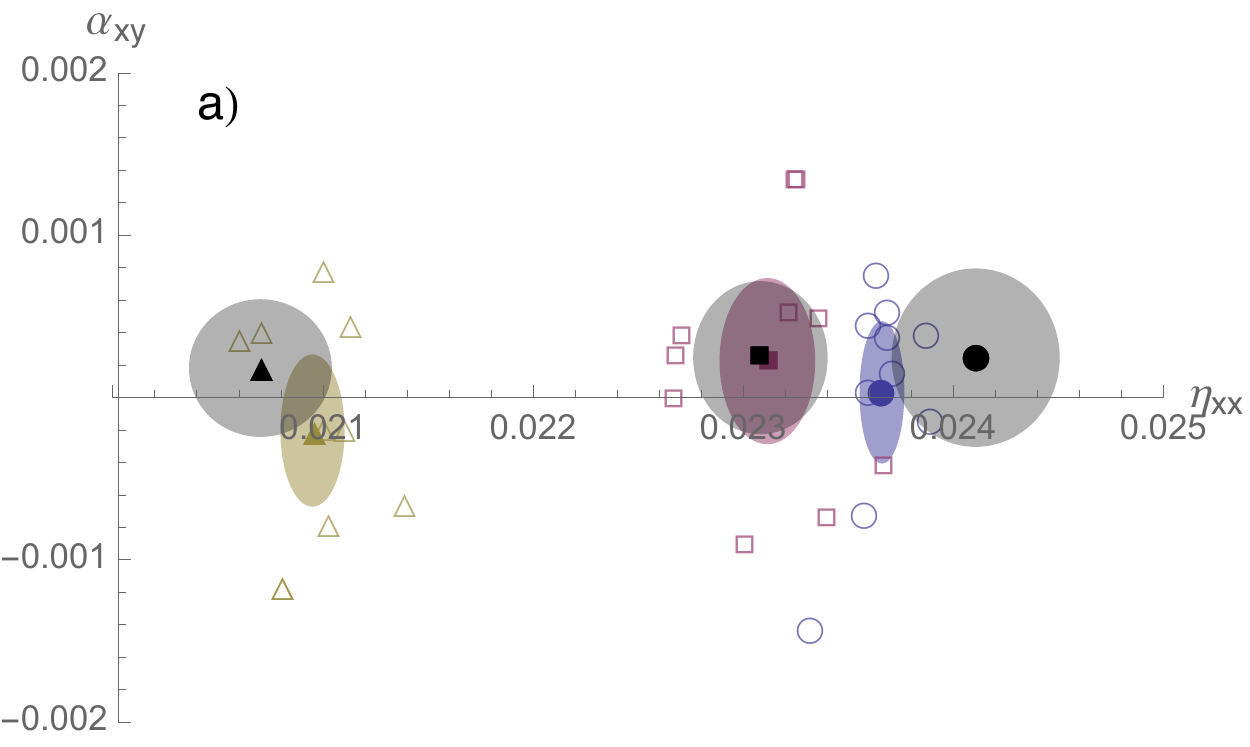}~~~~~\includegraphics[width=0.47\textwidth]{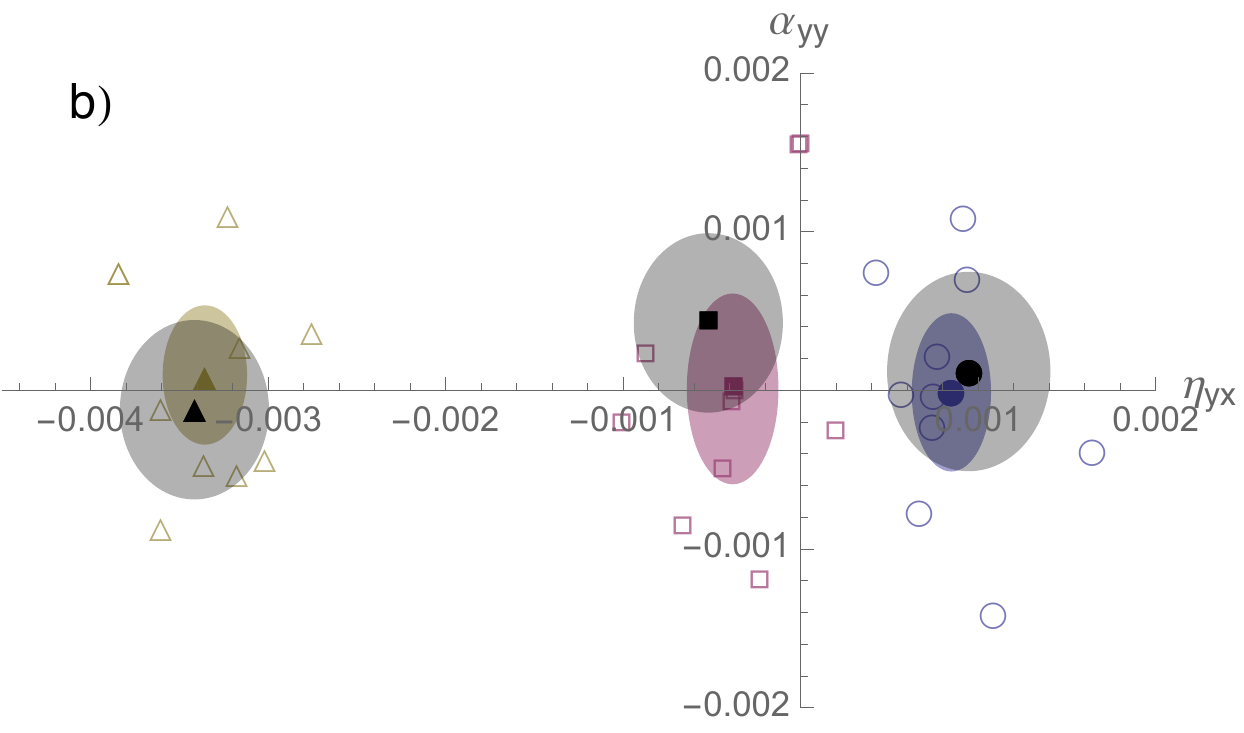}

\caption{Measured transport coefficients for $\mathrm{Re}=\mathrm{Rm}=100$
shearing-box simulations (as described in the text). (a) $\eta_{xx}$,
(b) $\eta_{yx}$. Squares, circles and triangles show $\Omega=0$,
$\Omega=4/3$ and $\Omega=4$ respectively, with the hollow markers
illustrating those measured from each simulation. The mean of these
measurements is shown by the solid colored marker, with its error
shown with the shaded circle (calculated from the standard deviation).
Test-field method results, against which least-squares results should
be compared, are illustrated by black markers, with the shaded area
showing the error in these measurements. (We have also included the
measured $\alpha$ values here, although in all cases these are zero
to within error.)\label{c5:fig:Measured-transport-coefficients}}
\end{centering}
\end{figure}

Test-field calculations are conducted as discussed in Sec.~\ref{sec:eta measurements}. Due to the lack of a
small-scale dynamo, the $\bm{b}$ fluctuations quickly reach a steady
state, and an average of $\bm{\mathcal{E}}$ is taken over $t=0\rightarrow1600$
to obtain $\eta_{xx}$ and $\eta_{yx}$. Errors are obtained
through the standard deviation of the mean after dividing 
the data set into $100$ bins  ($\pm$ values indicate the $95\%$ confidence interval). 
Results from the test-field method, which we consider as the reference
values against which to compare coefficients obtained using the fitting
method, are illustrated in Fig.~\ref{c5:fig:Measured-transport-coefficients} in black. 
These values are comparable (in the ratio of $\eta_{xx}$ to $\eta_{yx}$)
to those obtained in previous work for the non-rotating case \citep{Brandenburg:2008bc},
as well as exhibiting the expected trends (\citealt{Radler:2006hy}; \citetalias{LowRm}).
In all cases the test-field measured $\alpha_{ij}$ are zero to within
error  (see Fig.~\ref{c5:fig:Measured-transport-coefficients}). 

Results obtained by using the projection method on the self-consistent MHD
simulations are also illustrated
in Fig.~\ref{c5:fig:Measured-transport-coefficients}. The
transport coefficients $\left(\alpha_{xy},\alpha_{yy},\eta_{xx},\eta_{yx}\right)$ are measured
as described in Sec.~\ref{sec:eta measurements} for the duration of each simulation,
excluding times after which the dynamo has saturated. Because of the
long averaging time in comparison to the $\mathrm{Rm}=2000$ measurements presented in
Sec.~\ref{sec:numerical results}, the spread of values between different simulations
is quite small. 
It is seen that the measured coefficients agree with the test-field
calculations to within error margins in all cases. The largest discrepancy
is in $\eta_{xx}$ at $\Omega=0$, which may be related to the vorticity
dynamo (i.e., mean flow generation) that develops without rotation (the difference is still only
of the order of $1\%$). Note that a growing dynamo is observed in
\emph{all} of the self-consistent simulations, and at $\Omega=0$ this is
purely due to a stochastic-$\alpha$ effect (as discussed in \citetalias{LowRm}),
since the measured transport coefficients indicate the dynamo should be stable.
We can thus be sure that the fitting method
is not somehow measuring a property of the dynamo growth rate rather
than coherent transport coefficients.


\end{document}